\documentclass[fleqn]{2017SCGE}
\usepackage{multirow}
\usepackage{graphicx}
\usepackage{subfigure}

\begin{document}

\ensubject{subject}

\ArticleType{Article}
\SpecialTopic{SPECIAL TOPIC: }
\Year{0000}
\Month{00}
\Vol{00}
\No{0}
\DOI{xxxxxxxxxx}
\ArtNo{000000}
\ReceiveDate{00, 00, 0000}
\AcceptDate{000, 00, 0000}

\title{Holographic images  of a charged black hole in Lorentz symmetry breaking massive gravity}

\author[1]{Xiao-Xiong Zeng}{}
\author[*2]{Li-Fang Li}{}\thanks{Corresponding author. lilifang@imech.ac.cn}
\author[2]{Pan Li}{}
\author[2]{Bo Liang}{}
\author[2,3]{Peng Xu}{}



\AuthorMark{X. Zeng, L. Li, P. Li, B. Liang,  P. Xu}

\AuthorCitation{Xiao-Xiong Zeng, Li-Fang Li, Pan Li, Bo Liang, Peng Xu}

\address[1]{Department of Mechanics, Chongqing Jiaotong University, Chongqing 400074, China.}
\address[2]{Center for Gravitational Wave Experiment, National Microgravity Laboratory, Institute of Mechanics, Chinese Academy of Sciences, Beijing 100190, China.}
\address[3]{Lanzhou Center of Theoretical Physics, Lanzhou University, No. 222 South Tianshui Road, Lanzhou 730000, China.}


\abstract{Using the AdS/CFT correspondence, this paper investigates the holographic images of a charged black hole within the context of Lorentz symmetry breaking massive gravity. The  photon rings, luminosity-deformed rings, or light points from various observational perspectives are obtained. We also study  the influences of both the chemical potential and temperature  on the Einstein ring. Unlike the previous  work, which primarily examines the effect of chemical potential on ring radius at high temperatures and find no change in the radius with varying chemical potential, we also investigate the effect of chemical potential on the ring radius at low temperature besides at high temperature. Our findings indicate that at low temperatures, the photon ring radius decreases with increasing of chemical potential, while at high temperatures, the results are consistent with previous studies. Additionally, we explore the impact of the model parameter $\lambda$ on the Einstein ring radius and find the the ring radius increases as the model parameter $\lambda$ increases. More interestingly, for the large chemical potential, ${u} = 1$, the temperature dependence of the photon ring radius is reversed for $\lambda = 2$ and $\lambda = 4$. Conversely, for a small chemical potential ${u} = 0.1$, the temperature dependence of the Einstein ring stays the same as $\lambda = 2$ and $\lambda = 4$.}



\maketitle


\begin{multicols}{2}

\section{Introduction}
The General Theory of Relativity (GR), formulated in 1915, has been proved extensively to be validated on the Solar System and other astrophysical scales~\cite{Will2014}. It forms the foundation of the Standard Cosmological Model. However, certain observations and theoretical issues, such as the non-renormalization of GR, the cosmological constant problem, the late acceleration of the Universe~\cite{SP1999}, and anomalous galaxy rotation curves~\cite{Sofue2001}, remain unresolved by GR alone. In recent years, modified gravity theories have emerged as a broad class of models addressing these issues in the $\Lambda$CDM model and GR. These theories enhance cosmological evolution at both background and perturbation levels~\cite{Yashar2021}, including $f(R)$ gravity ~\cite{Cognola2008}, Gauss-Bonnet and $f(G)$ gravity~\cite{Antonio2009}, Lovelock gravity~\cite{Nathalie1990}, and others. Among these, massive gravity ~\cite{Fierz1939} is notable for potentially explaining many cosmological problems without invoking dark energy. There are many theories of massive gravity in the literature. In this paper we consider a massive gravity theory where the graviton acquires a mass due to the Lorentz symmetry breaking. Here Higgs mechanism for gravity is introduced with space-time depending scalar fields which are coupled to gravity. The resulting massive theory with Lorentz symmetry breaking exhibits modified gravitational interactions at large scale. And this theory is free from ghosts and tachyonic instabilities around the curved space as well as in the flat sapce. A more detailed review can be found in~\cite{Dubovsky:2004sg, Rubakov:2008nh}. This paper  focus on the massive gravity with spontaneously broken Lorentz symmetry.


Black shadows, representing the appearance of photons orbiting in unstable orbits around a black hole in an observer's sky~\cite{Ste2021, Ayz2018, Ku2022, An2023,Zeng:2020dco,Zeng:2020vsj}, offer a way to test modified gravity theories. The study of black holes and modern gravity has obtained interest not only from scientists but also from the general public, especially since the observation of Einstein rings~\cite{EventHorizonTelescope:2019dse, EventHorizonTelescope:2022wkp}. Specifically, an Einstein ring is formed by the deflection of photons emitted by light sources in the vicinity of a black hole~\cite{He2020}. When light rays approach a black hole, photons with low orbital angular momentum are captured by the black hole, creating a dark zone in the sky that can be observed by distant observers. This phenomenon is well-known as the Einstein ring~\cite{Falcke:1999pj, Lu2014, Dexter2014, Broderick2014, Bromley2001, Noble2007}. Following the first observation of the Einstein ring, there has been a surge in theoretical modeling efforts~\cite{Bambi2019, Vagnozzi2019, Allahyari2020, Khodadi2020, Babar2020, Kumar2020} to study colossal galaxies.

In the papers by Hashimoto et al.~\cite{Hashimoto:2018okj, Hashimoto:2019jmw}, a holographic method was proposed to study the Einstein ring using the AdS/CFT correspondence. This holographic dual~\cite{Witten1998, Horatiu2015, AmmonandJ2015} relates a gravity theory, typically in a $(d+1)$-dimensional Anti-de Sitter (AdS) background, to a conformal field theory (CFT) in $d$ dimensions at the boundary of the space. This has significantly advanced our understanding of string theory and quantum gravity. In~\cite{Hashimoto:2018okj, Hashimoto:2019jmw}, Hashimoto and colleagues considered a quantum field theory (QFT) on $R_t \times S^2$ and introduced an external source on $S^2$. Their results demonstrated that the well-known Einstein rings could be reproduced using the holographic method. Building on this pioneering work, Liu et al.~\cite{Liu:2022cev} investigated the holographic image of a charged AdS black hole using this approach. Their results indicated that the radius of the ring in the image is affected by temperature but not by the chemical potential. Further studies on the holographic images of black holes have extended  to various gravitational frameworks~\cite{Zeng:2023zlf, Zeng:2023tjb, Hu:2023eow, Zeng:2023ihy, Hu:2023mai,Zeng:2024ybj,Gui:2024owz}.

Although the overarching concept is similar~\cite{Zeng:2023zlf, Zeng:2023tjb, Hu:2023eow, Zeng:2023ihy, Hu:2023mai,Zeng:2024ybj,Gui:2024owz}, there are significant differences in the details. For example, the photon sphere's characteristics vary according to the specific bulk dual geometry. In this paper, we investigate the behavior of the lensed response in a charged black hole within the context of Lorentz symmetry breaking massive gravity. First, we study the impact of the frequency $\omega$ on the Einstein ring, finding that higher frequencies have a diminishing effect on the ring. We then focus on the effect of the chemical potential $u$ on the Einstein ring's radius at both high and low temperatures. Our results show that at low temperatures, the Einstein ring radius decreases with increasing chemical potential, whereas at high temperatures, the results are consistent with the previous studies. We also examine the influence of the model parameter $\lambda$ on the Einstein ring radius for $\lambda = 2$ and $\lambda = 4$. When $\lambda = 2$, the charged black hole reverts to the Reissner-Nordström AdS (RN-AdS) black hole. For a chemical potential $u = 1$, the temperature dependence behavior of the photon ring radius is reversed between $\lambda = 2$ and $\lambda = 4$. For a small chemical potential, such as $u = 0.1$, the temperature dependence behavior of the Einstein ring is consistent for both $\lambda = 2$ and $\lambda = 4$.

At present, there are indeed some works to study Einstein rings of black holes through holography. Especially in our previous work [33], we studied the effect of chemical potential on the ring where only the effect of chemical potential on the ring at high temperatures is considered. Here we further considered the case of low temperatures and found that the effect of chemical potential on the ring at low temperatures is completely different. On the other hand, we chose a Lorentz symmetry breaking massive gravity model, which can become to RN black holes under certain case. We also discuss the effects of temperature and chemical potential on the rings for different $\lambda$ at high temperature and low temperature with large chemical potential and small chemical potential. In this paper, the influence of chemical potential and temperature on the Einstein rings are more comprehensively and profoundly.

This paper is organized as follows. In Sec.~\ref{Introduction}, we introduce the charged black hole solution in Lorentz symmetry breaking massive gravity. In Sec.~\ref{construction}, we construct the holographic images and analyze the effects of the frequency $\omega$ and the chemical potential $u$ on the Einstein rings. In Sec.~\ref{3}, we compare the results obtained from the holographic method with those derived from geometric optics, demonstrating that the position of the photon ring obtained from geometric optics is in full agreement with that of the holographic ring.

\section{The charged black hole solution in Lorentz symmetry breaking in massive gravity}
\label{Introduction}
As we know, in 1939, massive gravity was first proposed by Fierz and Pauli~\cite{Fierz1939}. Here, we consider the Lorentz symmetry breaking massive gravity~\cite{Dubovsky:2004sg, Rubakov:2008nh}. The action of this Lorentz symmetry breaking massive gravity is given by
\begin{equation}
S=\int d^4 x \sqrt{-g} \left[ -M_{pl}^2 \mathcal{R}+\Omega^4 \mathcal{F}(X,W^{ij}) \right],
\end{equation}
here $\mathcal{R}$ is the scalar curvature and the function $\mathcal{F}$ describes the scalar fields $\Xi^0$ and $\Xi^{i}$. The scalar field is minimally coupled to gravity while in the theory the Lorentz symmetry is spontaneously broken. \(M_{pl}\) is the Planck mass and the parameter $\Omega$ is proportional to the graviton mass. It is possible to express the function $\mathcal{F}$ in terms of two Goldstone fields \(X\) and \(W^{ij}\). The expressions for \(X\) and \(W^{ij}\) are given as~\cite{Liang:2017ceh, Fernando:2018fpq, Fernando:2016qhq, Liu:2017jbm}
\begin{eqnarray}
X=\frac{\partial^{\rho}\Xi^0 \partial_{\rho}\Xi^{0}}{\Omega^4},
\end{eqnarray}
and 
\begin{equation}
W^{ij}=\frac{\partial^{\rho}\Xi^i \partial_{\rho}\Xi^{j}}{\Omega^4}-\frac{\partial^{\rho}\Xi^i \partial_{\rho}\Xi^{0}\partial^{\sigma}\Xi^{j}\partial_{\sigma}\Xi^0}{\Omega^4}.
\end{equation}
Finding analytical solutions for a generic function $\mathcal{F}$ is not possible. We can choose this function in such a way that the resulting equations are solvable analytically. We consider a static and spherically symmetric metric~\cite{Liang:2017ceh, Fernando:2018fpq, Fernando:2016qhq, Liu:2017jbm}
\begin{equation}
ds^2=-G(r)dt^2+K(r)dr^2+r^2 (d\psi^2+\sin^2 \psi d\xi^2)
\end{equation}
with
\begin{eqnarray}
\Xi^{0}=\Omega^2(t+N(r)), \ \ \ \Xi^{i}=\xi(r)\frac{\Omega^2 x^i}{r},
\end{eqnarray}
along with the choice~\cite{Liang:2017ceh, Fernando:2018fpq, Fernando:2016qhq, Liu:2017jbm} 
\begin{equation}
\mathcal{F}=c_0 (\frac{1}{X}+w_{1})+c_1(w_1^3-2w_1w_2-6w_1+2w_3-12),
\end{equation}
where $c_0$ and $c_1$ are some dimensionless constants and 
\begin{eqnarray}
w_1=-(f_1+2f_2), w_2=f_1^2+2f_2^2,w_3=-(f_1^3+2f_2^3),   
\end{eqnarray}
with
\begin{eqnarray}
f_1=\frac{\xi^{\prime 2}}{GKX^{\prime}}, \ f_2=\frac{\xi^2}{r^2},\ X=\frac{K-G N^{\prime 2}}{GK}.
\end{eqnarray}
With the above \(\mathcal{F}\), a static spherically symmetric metric solution of a charged AdS black hole is as follows~\cite{Bebronne:2009mz}
\begin{equation}
ds^2=-G(r)dt^2+\frac{1}{G(r)}dr^2+r^2(d\psi^2+\sin^2\psi d\xi^2), \label{metric}
\end{equation}
and the corresponding metric function is
\begin{equation}
G(r)=1-\frac{2M}{r}-\chi\frac{Q^2}{r^{\lambda}}-\frac{\Lambda r^2}{3},
\label{G}
\end{equation}
where \(\Lambda\) is the cosmological constant. For more details, please refer to the ~\cite{Liang:2017ceh, Fernando:2018fpq, Fernando:2016qhq, Liu:2017jbm}. 
At the event horizon, the Hawking temperature is calculated as follows
\begin{equation}
T=\frac{f^{\prime}}{4\pi}=\frac{1}{4\pi} \left( \frac{1}{r_h}-r_h\Lambda+\frac{\chi (\lambda-1)}{r_h^{\lambda+1}}Q^2 \right).\label{tr}
\end{equation}

Here, we carefully analyze the two parameters $\chi$ and $\lambda$ in Eq. (\ref{G}). For $\chi=-1$, this metric solution is similar to the well-known Reissner-Nordström AdS black hole, which admits a nontrivial small or large black hole phase transition~\cite{Fernando:2016qhq}. Therefore, we choose $\chi=-1$ in this paper in order to compare our results with the case of Reissner-Nordström AdS black hole. For the parameter $\lambda$, the ADM mass will diverge if $\lambda<1$, which is not allowed~\cite{Fernando:2016sps}, so we let $\lambda>1$. 
To sum up, in this paper we consider the case for $\chi=-1$ and $\lambda>1$.

Next, we focus on the spherical metric (\ref{G}). With a new definition \(r_*=1/r\), we have \(F(r)=r_*^{-2}F(r_*)\). We rewrite the metric (\ref{G}) in the new coordinates \((t, r_*, \theta, \phi)\) and derive
\begin{equation}
ds^2=\frac{1}{r_*^2}\left[-F(r_*)dt^2+\frac{d r_*^2}{F(r_*)}+d\Omega^2\right],
\label{metric_2}
\end{equation}
where \(d\Omega^2=d\theta^2+\sin^2\theta d\phi^2\). We will take Eq. (\ref{metric_2}) as our background next.
The complex scalar field \(\Phi\) in such a background is determined by the Klein-Gordon equation~\cite{Liu:2022cev}
\begin{equation}
D_b D^b \Phi - M^2\Phi=0,
\end{equation}
where \(D_a=\nabla_a - i e A_a\) and we will take \(M^2=-2\) in this paper. It is worth pointing out that we consider the weak coupling problem and only the effect of the metric on the scalar field is involved, meaning the back reaction of the scalar field on the metric is not included. To solve the above Klein-Gordon equation easily, the ingoing Eddington coordinate is introduced, which is expressed as
\begin{equation}
\nu=t-\int \frac{d r_*}{F(r_*)}.
\end{equation}
With the above new coordinate, the metric (\ref{metric_2}) is further expressed as
\begin{equation}
ds^2=\frac{1}{r_*^2}\left[-F(r_*)d\nu^2-2 d r_* d\nu+d\Omega^2\right].
\end{equation}

The corresponding asymptotic solution of this complex scalar field \(\Phi\) near the AdS boundary is~\cite{Liu:2022cev}
\begin{equation}
\Phi(\nu,r_*,\theta,\phi)=r_* J_{\mathcal{O}}(\nu,\theta,\phi)+r_*^2 \langle \mathcal O \rangle + \mathrm{O}(r_*^3).
\end{equation}
Based on the AdS/CFT correspondence, \(J_{\mathcal{O}}\) is the source for the boundary field theory, and its expectation value is
\begin{equation}
\langle \mathcal O \rangle_{J_{\mathcal{O}}} = \langle \mathcal O \rangle - (\partial_{\nu} - ie u) J_{\mathcal{O}},
\end{equation}
where \(\langle \mathcal{O} \rangle_{J_{\mathcal{O}}}\) is also called the response function, and \(u= Q/r_h\) is the chemical potential. It is evident that \(\langle \mathcal{O} \rangle\) corresponds to the expectation value of the dual operator when the source is turned off.

\section{Holographic images construction of the charged black hole in Lorentz symmetry breaking massive gravity}
\label{construction}

\begin{figure}[H]
\centering
\includegraphics[scale=0.55]{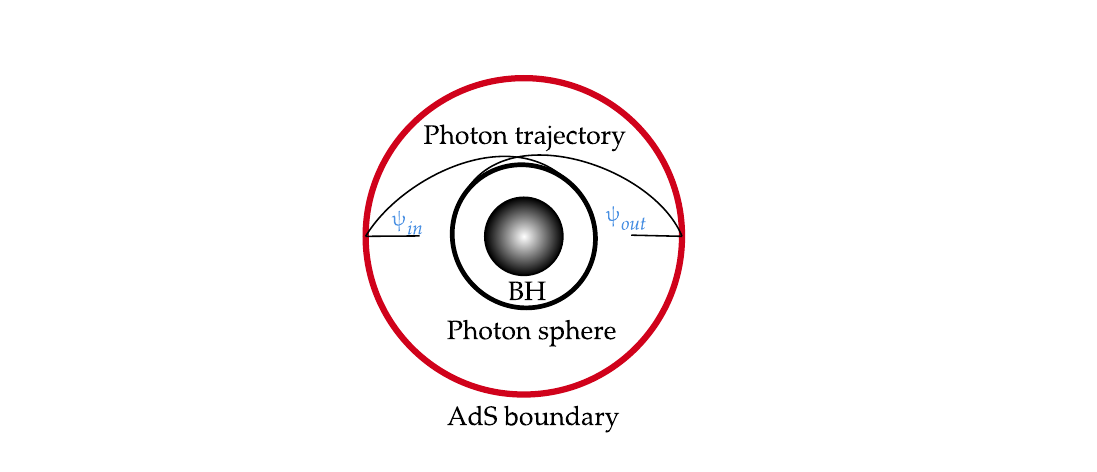}
 \caption{The ingoing angle and outgoing angle corresponding to the moving of the photons around the black hole. } \label{go_es}
\end{figure}

In 2019, the Event Horizon Telescope (EHT) collaboration claimed the first capture of a shadow image of the supermassive black hole at the center of the galaxy M87~\cite{EventHorizonTelescope:2022wkp}. This breakthrough has sparked increased scientific interest in black hole shadow research. In this section, we introduce the holographic construction of Einstein rings for charged black holes in Lorentz symmetry-breaking massive gravity. Similar to previous holographic studies, the external source is mapped to the boundary condition of the bulk field. A probe wave emitted from the source located on the AdS boundary is diffracted by the black hole as it traverses through the black hole spacetime, ultimately reaching the opposite side of the AdS boundary, as depicted in Fig.~\ref{go_es}.

Considering the time-periodic boundary conditions, we select  an axially symmetric, monochromatic oscillating Gaussian source, positioned at the south pole of the AdS boundary. Its explicit expression is as follows
\begin{eqnarray}
J_{\mathcal{O}}(\nu_e,\psi) &=& \frac{1}{2\pi \sigma^2} e^{-i \bar{\omega} \nu_e} e^{-\frac{(\pi-\psi)^2}{2\sigma^2}} \nonumber \\
&=& e^{-\bar{\omega}\nu_e} \sum_{l=0}^{\infty} C_{l0} Y_{l0}(\psi),
\end{eqnarray}
where $\nu_e$ and $\psi$ are coordinates, and $\sigma$ is the width of the wave. When $\sigma \ll 1$, the Gaussian tail is negligible, and the Gaussian function decomposes into scalar spherical harmonics. Thus, the source can be further expressed as
\begin{equation}
J_{\mathcal{O}}(\nu_e,\psi) = e^{-i \omega \nu_e} \sum_{l=0}^{\infty} C_{l0} Y_{l0}(\psi).
\end{equation}
The coefficients $C_{l0}$ preceding the spherical harmonics $Y_{l0}$ are given by
\begin{eqnarray}
C_{l0} &=& (-1)^l \sqrt{\frac{l+1/2}{2\pi}} \, e^{-\frac{1}{2}(l+1/2)^2 \sigma^2},
\end{eqnarray}
where $l$ denotes the magnetic quantum number, and $Y_{l0}$ represents the spherical harmonics function.

Considering spacetime symmetry, the complex scalar field $\Phi(\nu,r_*,\theta,\phi)$ is decomposed as
\begin{eqnarray}
\Phi(\nu,r_*,\theta,\phi) &=& e^{-i\omega \nu} \sum_{l=0}^{\infty} C_{l0} r_{*} \mathcal{Z}_l(r_*) Y_{l0}(\theta),
\end{eqnarray}
where $\mathcal{Z}_l$ satisfies the following equation of motion
\begin{eqnarray}\label{equation}
&&r_*^2 f \mathcal{Z}_l'' + r_*^2 [f' + 2 i (\omega - e A)] \mathcal{Z}_l' \nonumber \\
&+&  [(2-2f) + r_* f' - r_*^2 l(l+1) - i e r_*^2 A'] \mathcal{Z}_l = 0,
\end{eqnarray}
with $e$ denoting the charge of the complex scalar field. Near the AdS boundary, $\mathcal{Z}_l$ behaves as
\begin{equation}
\mathcal{Z}_l = 1 + r_* \langle \mathcal{O}_l \rangle + O(r_*^2).
\end{equation}
Based on this analysis, the response function $\langle\mathcal{O}\rangle_{J_{\mathcal{O}}}$ is expressed as
\begin{equation}
\langle\mathcal{O}\rangle_{J_{\mathcal{O}}} = e^{-i\omega \nu} \sum_{l=0}^{\infty} C_{l0} \langle \mathcal{O}\rangle_{J_{\mathcal{O}l}} Y_{l0}(\theta).
\end{equation}
Here, the response function $\langle \mathcal{O} \rangle_{J_{\mathcal{O}l}}$ is given by
\begin{equation}\label{response}
\langle \mathcal{O} \rangle_{J_{\mathcal{O}l}} = \langle \mathcal{O} \rangle_l + i \hat{\omega},
\end{equation}
where $\hat{\omega} = \omega + e u$. The diffracted wave by the black hole propagates through the spacetime and reaches another point on the AdS boundary, yielding a corresponding response function $\langle O(X) \rangle$. This lensed response function carries crucial information about the bulk spacetime geometry.

To derive the boundary response function, a comprehensive optical setup is required, including optical instruments, a convex lens, and a screen to capture the black hole image. With this setup, the response function can be extracted and analyzed in detail.

Using optical instruments, the response function $\langle O(X) \rangle$ is transformed into an image of the black hole on the screen $\Psi_{sc}(\hat{X}_{sc})$, satisfying
\begin{equation}\label{response_1}
\Psi_{sc}(\hat{X}_{sc}) = \int_{|\hat{x}| < {d}} d X^2 \langle \mathcal{O}(\hat{X}) \rangle e^{-i \frac{\bar{\omega}}{f} \hat{X} \cdot \hat{X}_{sc}},
\end{equation}
where $\hat{X} = (X,Y,Z)$ is the Cartesian-like coordinate on the boundary $S^2$, and $\hat{X}_{sc} = (X_{sc}, Y_{sc}, Z_{sc})$ represents a point on the screen. In this context, the convex lens acts as a converter, transforming the plane wave into a spherical wave that appears on the curved screen. Assuming the observer is positioned at $(\theta,\varphi) = (\theta_{obs},0)$, and the convex lens is adjusted on a three-dimensional flat surface such that $(X_{cl}, Y_{cl}) = (X,Y,0)$, the response function $\langle \mathcal{O}(X) \rangle$ projects as a planar wave denoted by $\Psi_{iw}(\hat{X})$, which is then transformed by the thin lens to $\Psi_{ow}(\hat{X})$ converging at the focal point $z=f$. Here, $D$ represents the propagation distance from the lens point $(X,Y,0)$ to the screen point $(X_{sc},Y_{sc},Z_{sc})$ and $d$ is the convex lens radius. This transformation relationship is expressed as
\begin{eqnarray}
\Psi_{ow}(\hat{X}) = e^{-i \bar{\omega} \frac{|\hat{X}|^2}{2f}} \Psi_{iw}(\hat{X}). \label{aa}
\end{eqnarray}
It is crucial that the focal length $f$ is significantly larger than the lens radius $d$. Finally, fixing the spherical screen at $(X,Y,Z) = (X_{sc}, Y_{sc}, Z_{sc})$ with the condition $X_{sc}^2 + Y_{sc}^2 + Z_{sc}^2 = f^2$, the wave equation $\Psi_{sc}(\hat{X}_{sc})$ is further expressed as
\begin{equation}
\Psi_{sc}(\hat{X}_{sc}) = \int_{|\hat{X}| \leq {d}} d^2 X \Psi_{ow}(\hat{X}) e^{-i \bar{\omega} D}. \label{ee}
\end{equation}
By inserting Eq. (\ref{aa}) and Eq. (\ref{ee}), we obtain
\begin{eqnarray}
\Psi_{sc}(\hat{X}_{sc}) &=& \int_{|\hat{X}| \leq {d}} d^2 X \Psi_{ow}(\hat{X}) e^{-i \bar{\omega} D} \nonumber \\
&=& \int_{|\hat{X}| \leq d} \Psi_{iw}(\hat{X}) e^{-i \frac{\bar{\omega}}{f} \hat{X} \cdot \hat{X}_{sc}} \nonumber \\
&=& \int_{|\hat{X}| \leq d} d^2 X \Psi_{iw}(\hat{X}) W(\hat{X}) e^{-i \frac{\bar{\omega}}{f} \hat{X} \cdot \hat{X}_{sc}},
\end{eqnarray}
where the window function $W(\hat{X})$ is defined as:
\begin{eqnarray}
W(\hat{X}) = \left\{
\begin{aligned}
0, & \quad 0 \leq |\hat{X}| \leq {d}, \\
1, & \quad |\hat{X}| \geq {d}.
\end{aligned}
\right.
\end{eqnarray}
\begin{figure}[H]
\centering
\subfigure
{\includegraphics[scale=0.36]{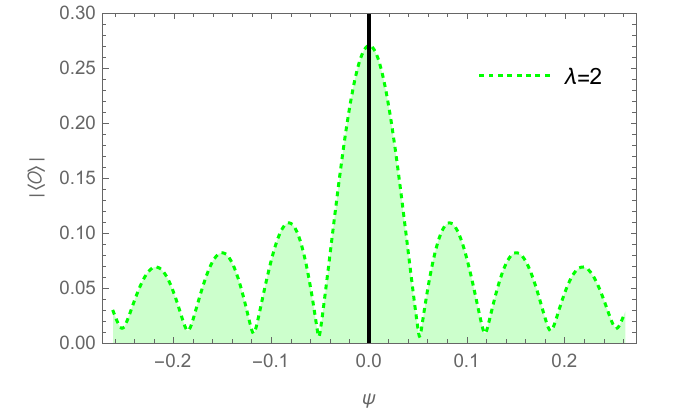}}
\subfigure
{\includegraphics[scale=0.36]{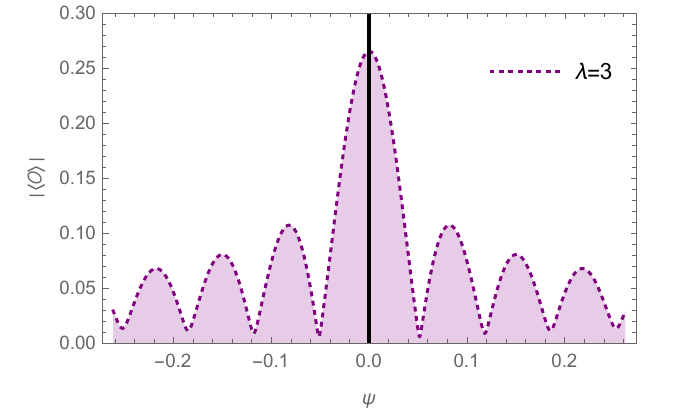}}\\
\subfigure
{\includegraphics[scale=0.36]{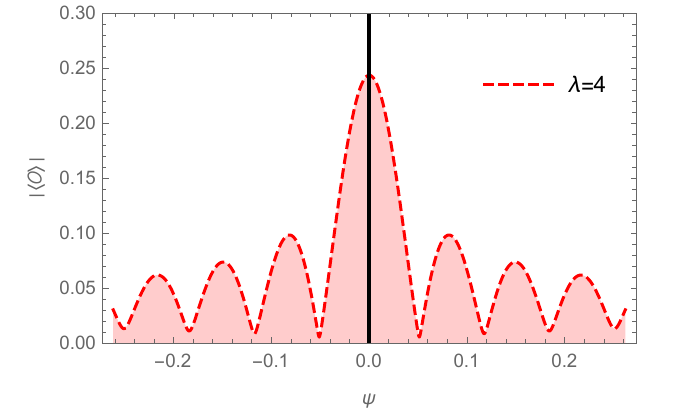}}
\subfigure
{\includegraphics[scale=0.36]{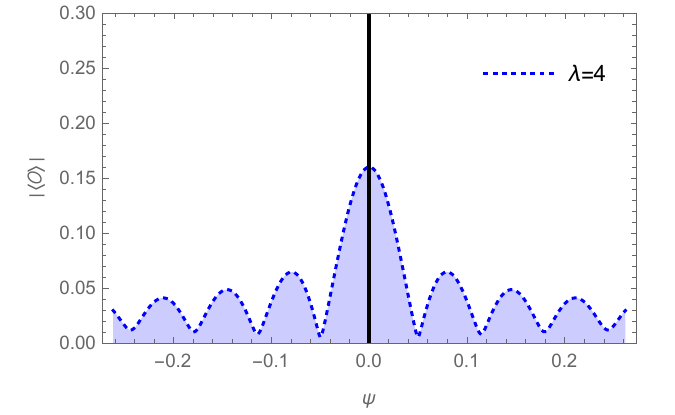}}
\caption{\label{fig1} Effect of different $\lambda$ on the response function, where $Q=0.01,z_h=4,\bar{\omega}=80,e=0.01$.}
\end{figure}
Our primary objective is to solve the radial Eq. (\ref{equation}) with the boundary condition $\mathcal{Z}_l(0)=1$ at the AdS boundary. Employing the pseudo-spectral method~\cite{Liu:2022cev}, we obtain the corresponding numerical solutions for $\mathcal{Z}_l$ and $\mathcal{O}_l$. With $\mathcal{O}_l$ extracted, the total response is obtained via Eq. (\ref{response}). We begin by plotting a typical profile of the total response $\langle \mathcal{O} \rangle$ in Fig.~\ref{fig1} to Fig.~\ref{fig_3}. All results confirm that the interference pattern arises from the diffraction of our scalar field by the black hole.

\begin{figure}[H]
\centering
\subfigure[$\lambda=2$]{
\includegraphics[scale=0.36]{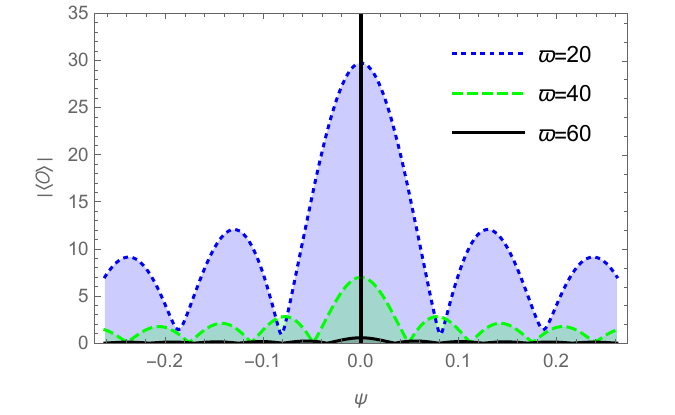}}
\subfigure[$\lambda=4$]{
\includegraphics[scale=0.36]{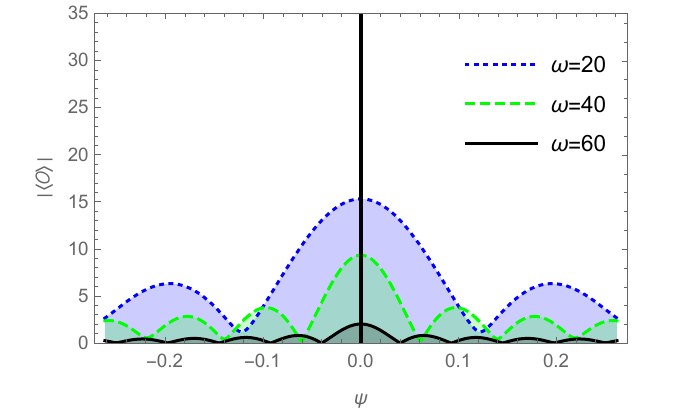}}
\caption{\label{fig2} Effect of frequency $\bar{\omega}$ on the response function when  $Q=1, z_h=1, e=0.01$.}
\end{figure}
We first exam the impact of the model parameter $\lambda$ on the response function for different values, specifically $\lambda=2$, $\lambda=3$, $\lambda=4$, and $\lambda=5$, while keeping $Q=0.01$, $z_h=4$, $e=0.01$, and $\bar{\omega}=80$, as depicted in Fig. \ref{fig1}. It is observed that varying $\lambda$ primarily affects the amplitude of the response function, while other characteristics remain nearly identical. 
We can see for $\lambda=2$, $\lambda=3$, $\lambda=4$, and $\lambda=5$, the  amplitude of the response function is about 0.27, 0.26, 0.24,0.16 respectively. In other words, as $\lambda$ increases, the magnitude of the response function decreases.  
 From Eq. \ref{response_1}, we will observe this change also showing up in the black hole image, which will be discussed detailed in next section.  

\begin{figure}[H]
\centering
\subfigure{
\includegraphics[scale=0.5]{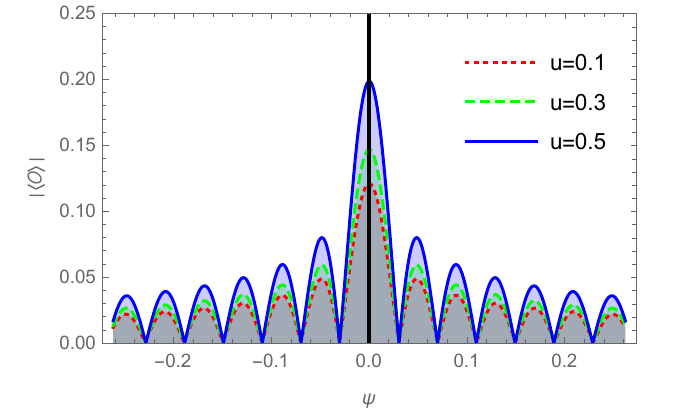}}
\caption{\label{fig_U} Effect of chemical potential  $u$ on the response function when $\lambda=4$, $\bar{\omega}=80,e=0.01,z_h=1$.}
\end{figure}

Moving to Fig. \ref{fig2}, we investigate the influence of the frequency $\bar{\omega}$ on the response function, focusing on $\lambda=2$ and $\lambda=4$ in $(\mathrm{a})$ and $(\mathrm{b})$ respectively, with parameters $Q=1$, $z_{h}=1$, and $e=0.01$. Here, we observe that the larger $\lambda$, the smaller the amplitudes of $|\langle O\rangle|$, as described in Fig. \ref{fig1}. Additionally, we note that increasing $\bar{\omega}$ shortens the oscillation period of the wave for both $\lambda=2$ and $\lambda=4$, indicating that $\bar{\omega}$ affects the width of the wave emitted by the Gaussian source. These findings highlight the critical dependence of the total response function on the properties of the Gaussian source. The same trend holds true for other values of $\lambda$ such as $\lambda=2,3,4$. It's noteworthy that when $\lambda=2$, the charged black hole in Lorentz symmetry breaking massive gravity reverts to the RN-AdS black hole discussed in ~\cite{Liu:2022cev}, in which the authors did not discuss the effect of  frequency $\bar{\omega}$ on the response function.

\begin{figure}[H]
\centering
\subfigure{
\includegraphics[scale=0.5]{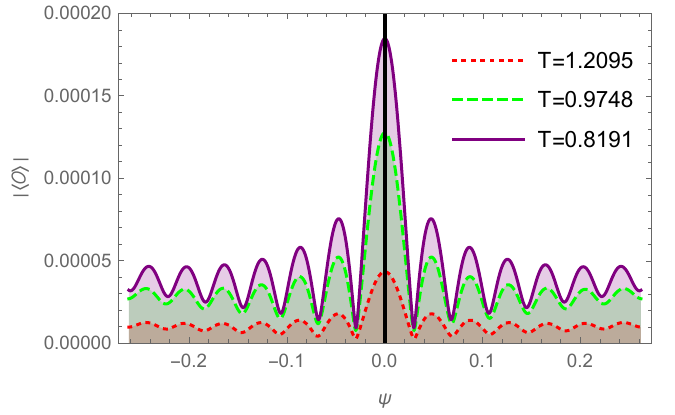}}
\caption{\label{fig_3} Effect of temperature  $T$ on the response function when $\lambda=4$, $\bar{\omega}=80,e=0.01, Q=1$.}
\end{figure}

We further analyze the amplitude of $\langle O \rangle$ under varying chemical potentials $u$, focusing on $\lambda=4$, $z_{h}=1$, $\bar{\omega}=80$, and $e=0.01$, shown in Fig. \ref{fig_U}. As $u$ increases from $0.1$ to $0.5$, the amplitude of $\langle O \rangle$ also increases gradually. This observation indicates a direct relationship between the chemical potential $u$ and the amplitude of the response function.
From Eq. \ref{response_1}, we see the change of the  chemical potential will also affect the images of the black hole.

\end{multicols}
\begin{figure}[H]
\centering
\subfigure[$\lambda=2,~\psi_{obs}=0^{o}$]{
\includegraphics[scale=0.33]{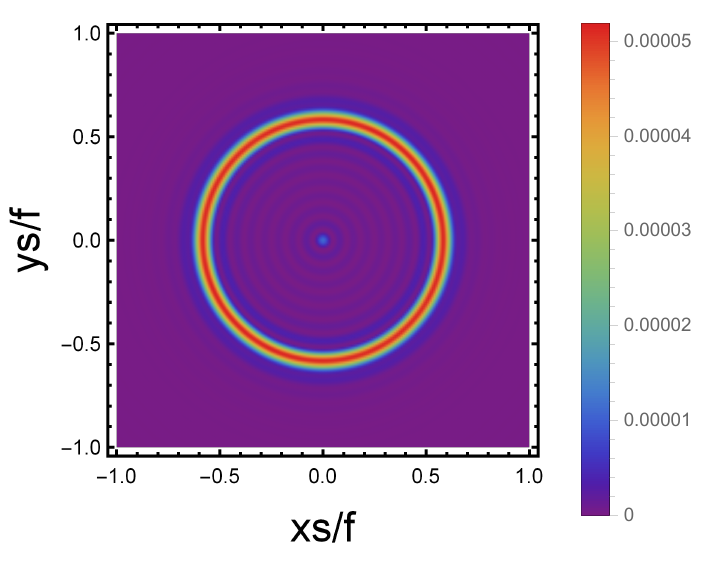}}
\subfigure[$\lambda=2,~\psi_{obs}=30^{o}$]{
\includegraphics[scale=0.33]{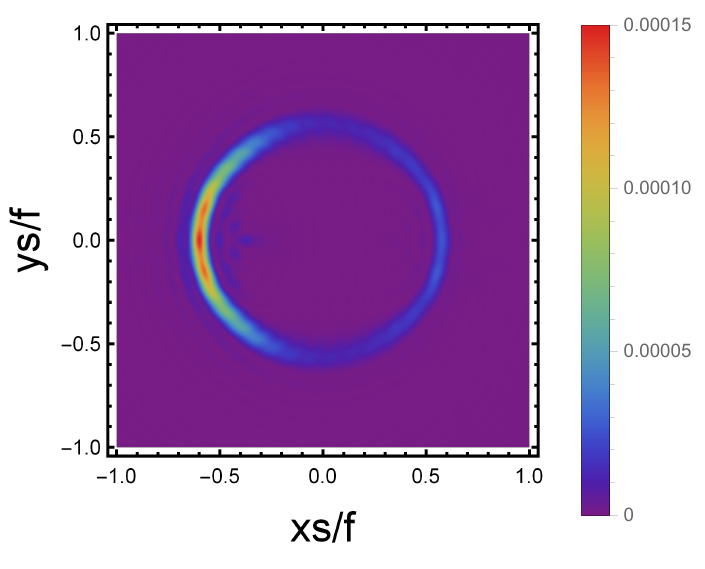}}
\subfigure[$\lambda=2,~\psi_{obs}=60^{o}$]{
\includegraphics[scale=0.33]{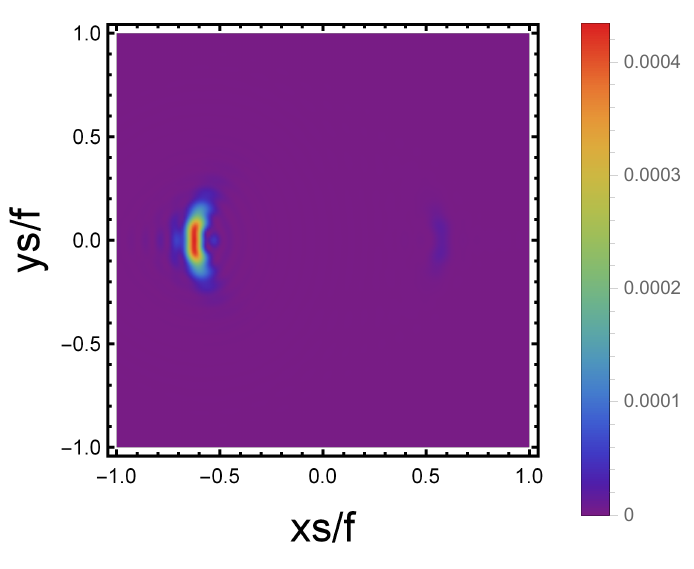}}
\subfigure[$\lambda=2,~\psi_{obs}=90^{o}$]{
\includegraphics[scale=0.33]{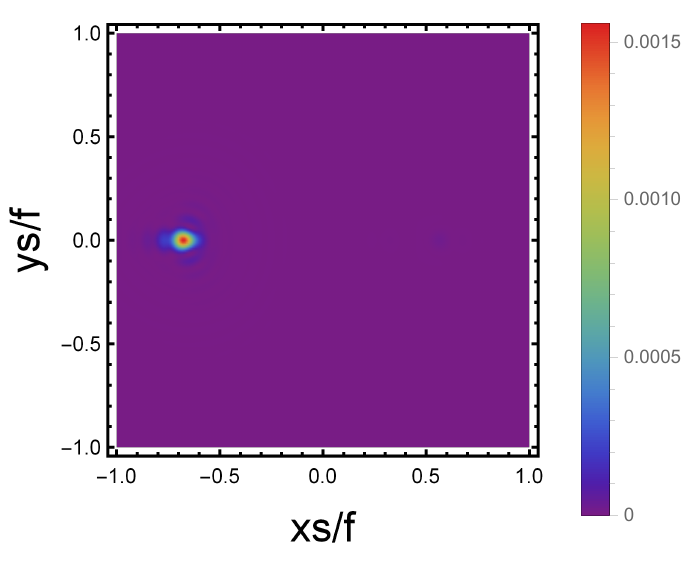}}
\subfigure[$\lambda=3,~\psi_{obs}=0^{o}$]{
\includegraphics[scale=0.33]{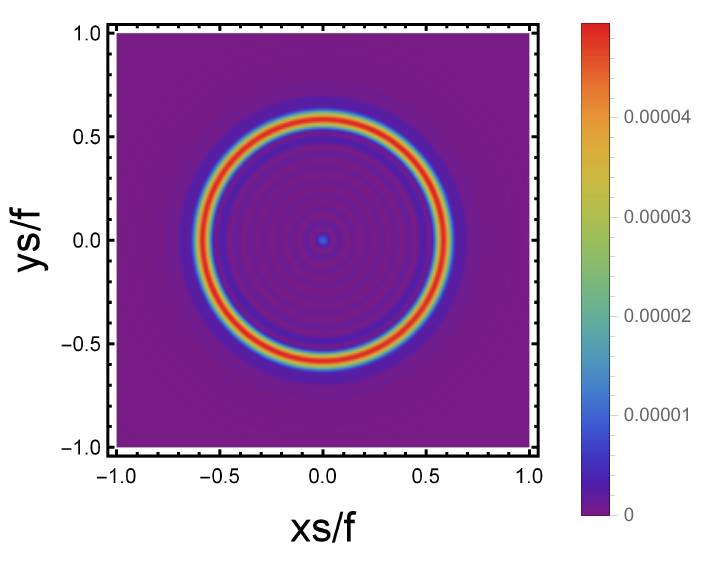}}
\subfigure[$\lambda=3,~\psi_{obs}=30^{o}$]{
\includegraphics[scale=0.33]{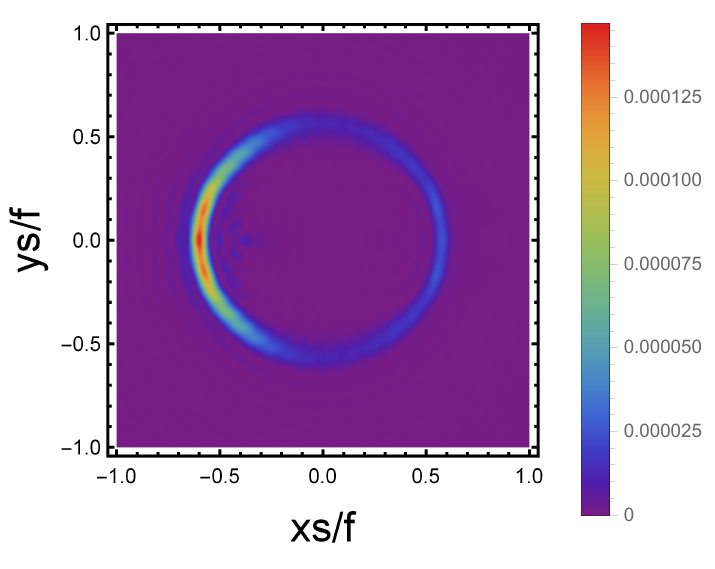}}
\subfigure[$\lambda=3,~\psi_{obs}=60^{o}$]{
\includegraphics[scale=0.33]{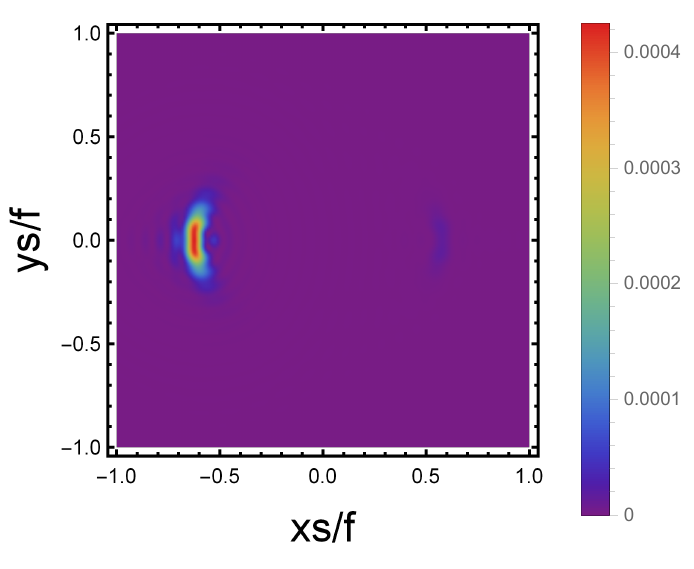}}
\subfigure[$\lambda=3,~\psi_{obs}=90^{o}$]{
\includegraphics[scale=0.33]{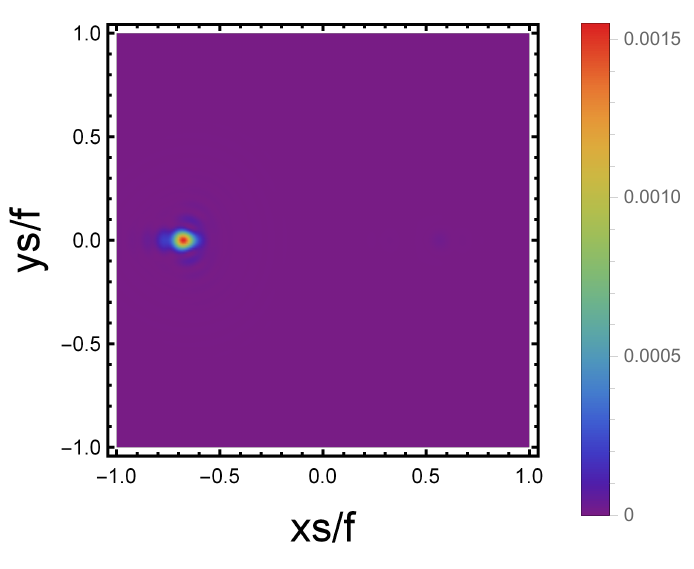}}
\subfigure[$\lambda=4,~\psi_{obs}=0^{o}$]{
\includegraphics[scale=0.33]{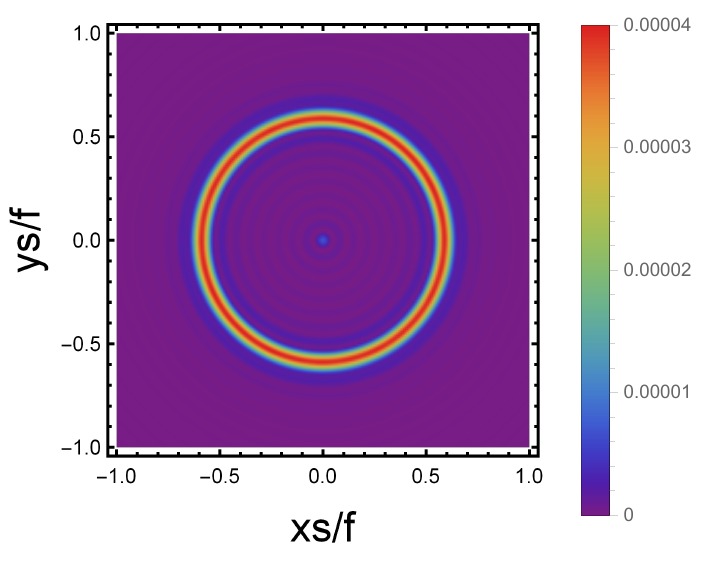}}
\subfigure[$\lambda=4,~\psi_{obs}=30^{o}$]{
\includegraphics[scale=0.33]{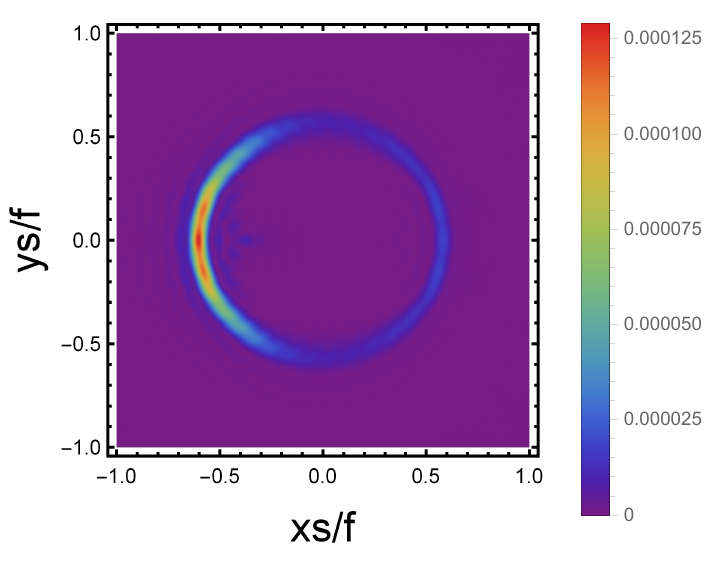}}
\subfigure[$\lambda=4,~\psi_{obs}=60^{o}$]{
\includegraphics[scale=0.33]{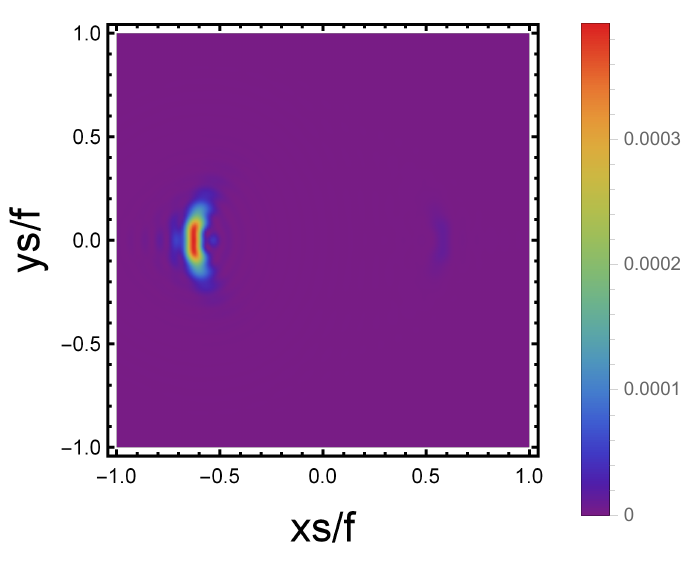}}
\subfigure[$\lambda=4,~\psi_{obs}=90^{o}$]{
\includegraphics[scale=0.33]{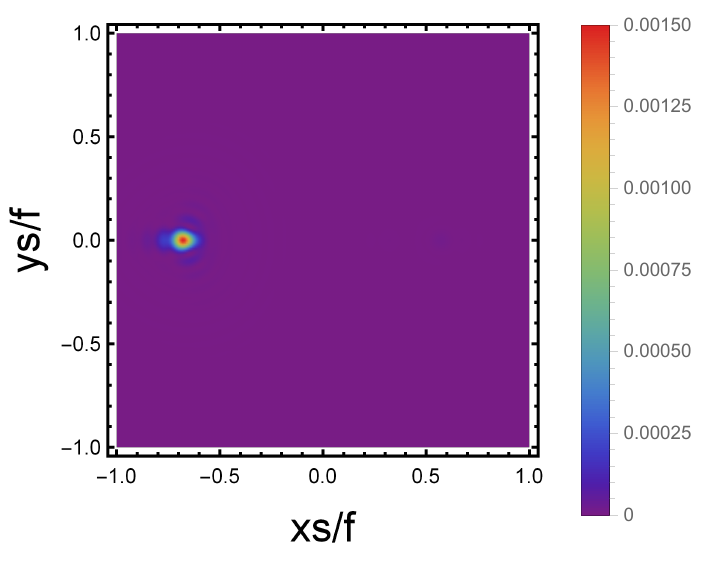}}
\subfigure[$\lambda=5,~\psi_{obs}=0^{o}$]{
\includegraphics[scale=0.33]{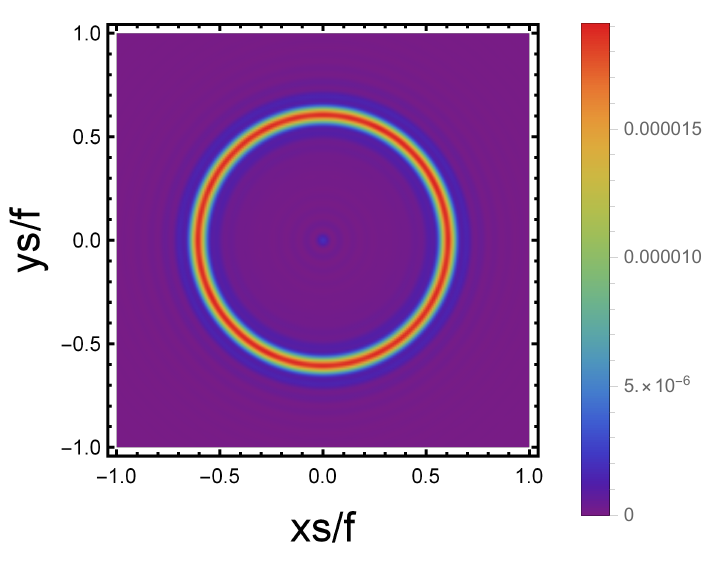}}
\subfigure[$\lambda=5,~\psi_{obs}=30^{o}$]{
\includegraphics[scale=0.33]{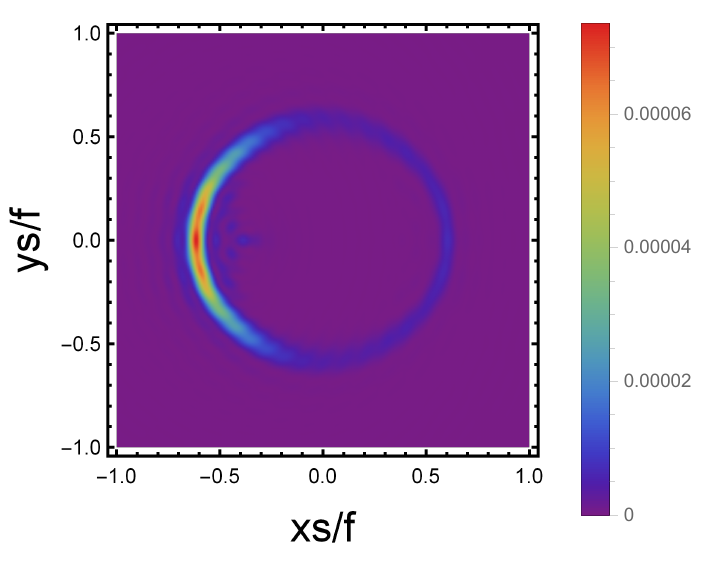}}
\subfigure[$\lambda=5,~\psi_{obs}=60^{o}$]{
\includegraphics[scale=0.33]{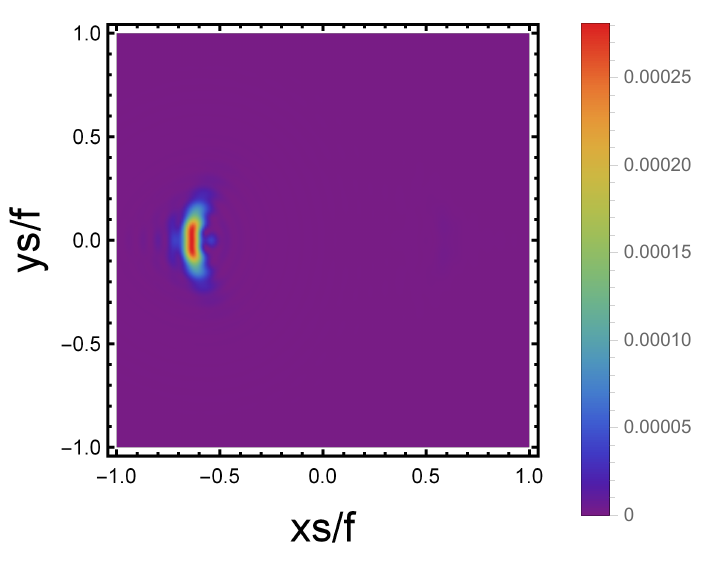}}
\subfigure[$\lambda=5,~\psi_{obs}=90^{o}$]{
\includegraphics[scale=0.33]{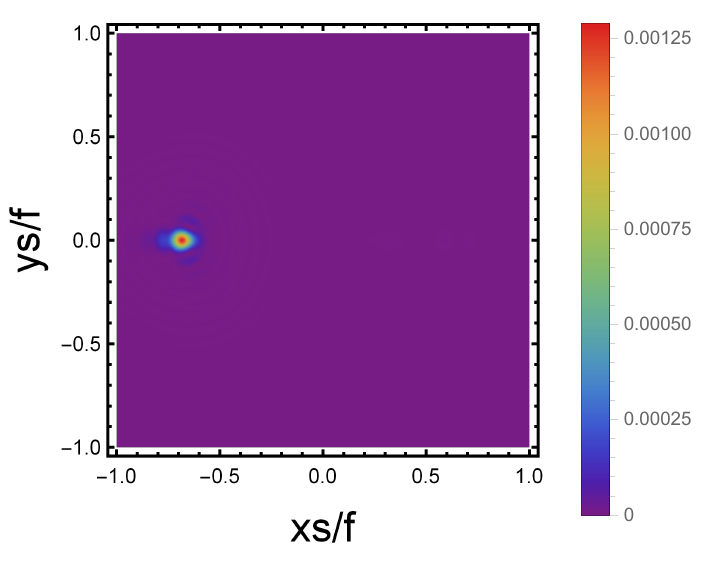}}
\caption{Observational appearance of the response on the screen for different $\lambda$ at various observation angles, where $Q=0.01,z_h=4,\bar{\omega}=80,e=0.01$. } \label{fig4}
\end{figure}

\begin{multicols}{2}

Next, we explore the amplitude of $\langle O \rangle$ across different temperatures $T$ in Fig. \ref{fig_3}, maintaining $\lambda=4$, $\bar{\omega}=80$, $e=0.01$ and $Q=1$. Here, we observe that as temperature increases, the amplitude of the total response function $\langle O \rangle$ decreases. This inverse relationship between temperature and amplitude implies the sensitivity of the response function to temperature variations.

Moving to Fig. \ref{fig4}, we present images of the lensed response observed at various observation angles for different $\lambda$, specifically with parameters $Q=0.01$, $z_{h}=4$, $\bar{\omega}=80$ and $e=0.01$. Each row corresponds to a fixed $\lambda$, but with a change of $\psi_{obs}$ from $0^{o}$ to $90^{o}$. Each columns correspond to a fixed $\psi_{obs}$ but with a change of the parameter $\lambda$ from $\lambda=2$ to $\lambda=5$. Therefore, from Fig. \ref{fig4}, we can see not only the effect of the observation angle on the black hole images, but also the effect of the model parameter $\lambda$  on the radius of the Einstein ring.   Notably, as $\psi_{obs}$ increases, the  spherically symmetric ring will be destroyed, especially when  $\psi_{obs}=90^{o}$, the Einstein ring will become a point.

\begin{figure}[H]
\centering
\subfigure[$\lambda=2$]{
\includegraphics[scale=0.42]{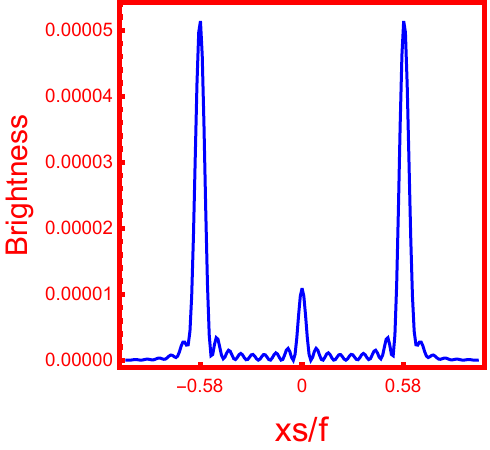}}
\subfigure[$\lambda=3$]{
\includegraphics[scale=0.42]{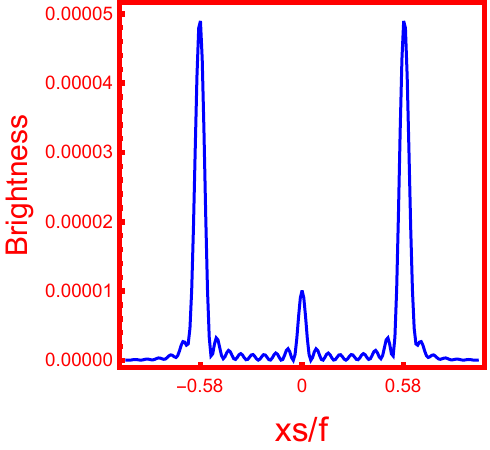}}
\subfigure[$\lambda=4$]{
\includegraphics[scale=0.42]{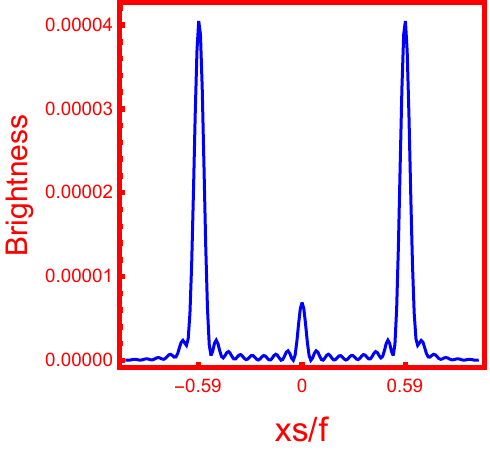}}
\subfigure[$\lambda=5$]{
\includegraphics[scale=0.42]{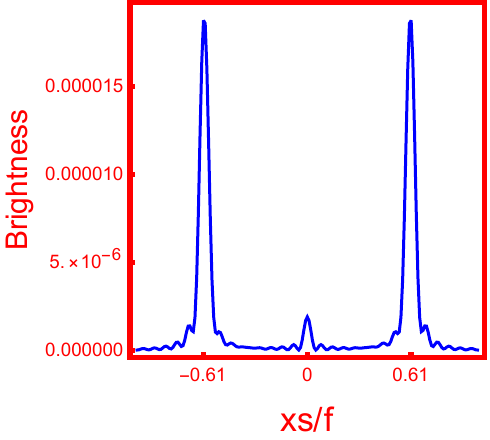}}
 \caption{The brightness  for different $\lambda$ at $\psi_{obs}=0^{o}$.} \label{fig55}
\end{figure}

Analyzing each column vertically, we find that the brightest of the Einstein ring decreases with the increase of the parameter $\lambda$, which is also clearly seen in  Fig. \ref{fig55}. In Fig. \ref{fig55}, the brightest value is 0.00005 for $\lambda=2$, while the brightest value is 0.000015 for $\lambda=5$. Revisit Fig. \ref{fig55}, we also find that as $\lambda$ increases, the radius of Einstein rings expands outward, that is to say, the concentric rings move further from the center. Although this change seems not  obvious in Fig. \ref{fig4}, we can clearly see  it in Fig. \ref{fig55}, which shows  brightness for different $\lambda$ at $\psi_{obs}=0$. The peak of these curves directly correlates with the radius of the rings. We can see that for $\lambda=2$, $\lambda=3$, $\lambda=4$, and $\lambda=5$, the peak locates at 0.58, 0.58, 0.59, 0.61 respectively, demonstrating that increasing $\lambda$ leads to larger ring radius.

\begin{figure}[H]
\centering
\subfigure[$\bar{\omega}=20$]{
\includegraphics[scale=0.35]{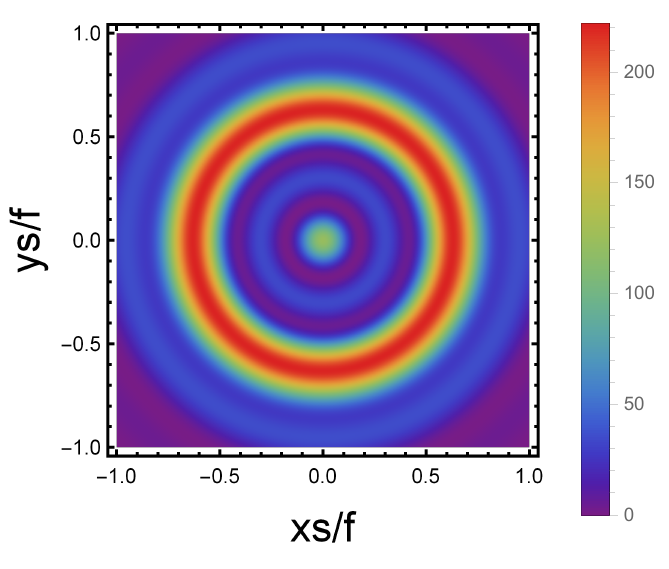}}
\subfigure[$\bar{\omega}=40$]{
\includegraphics[scale=0.35]{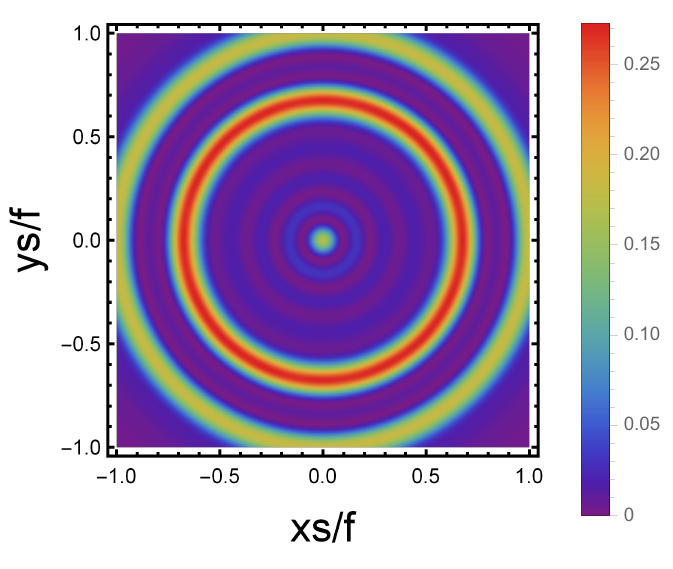}}
\subfigure[$\bar{\omega}=60$]{
\includegraphics[scale=0.35]{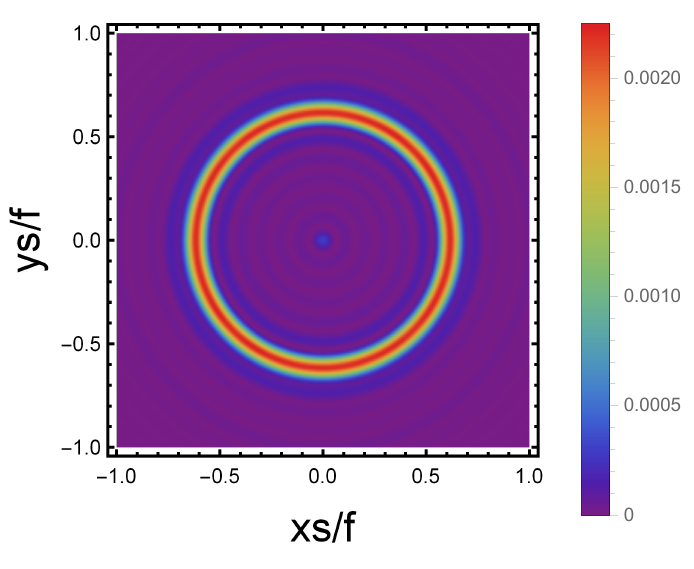}}
\subfigure[$\bar{\omega}=80$]{
\includegraphics[scale=0.35]{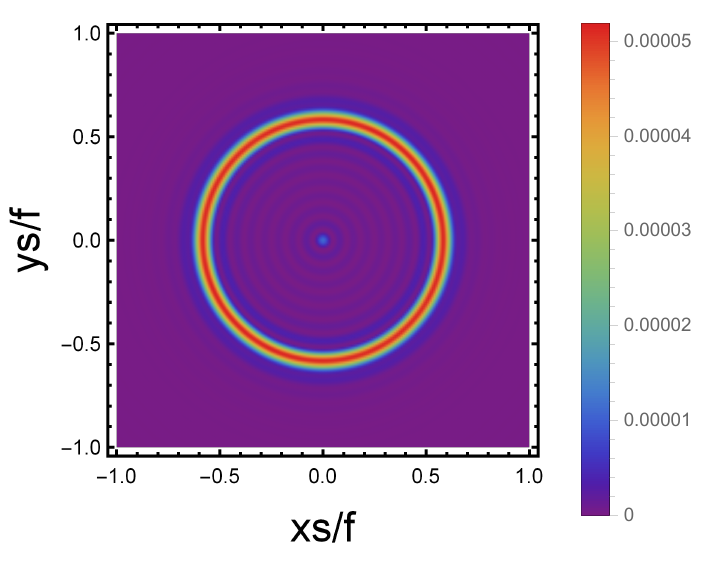}}
 \caption{The effect of  frequency $\bar{\omega}$ on the Einstein ring, when $Q=0.01,z_h=4,e=0.01$,$\lambda=2$.} \label{fig555}
\end{figure}

\begin{figure}[H]
\centering
\subfigure[$u=0.1$]{
\includegraphics[scale=0.33]{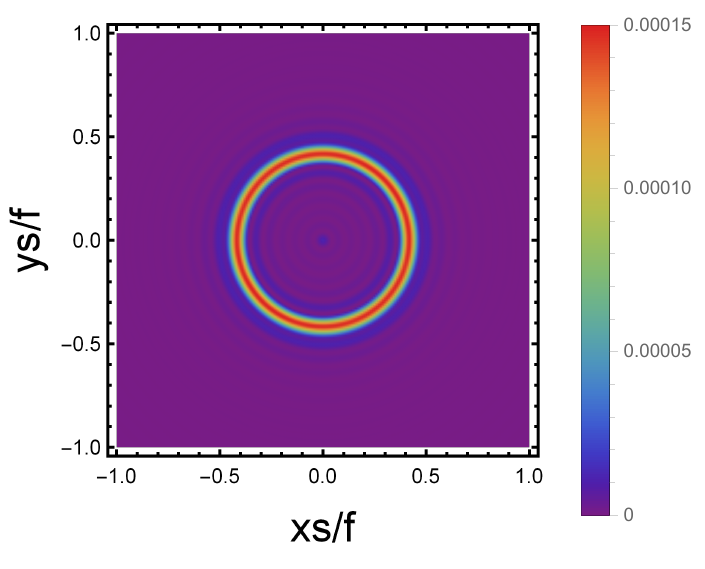}}
\subfigure[$u=0.3$]{
\includegraphics[scale=0.33]{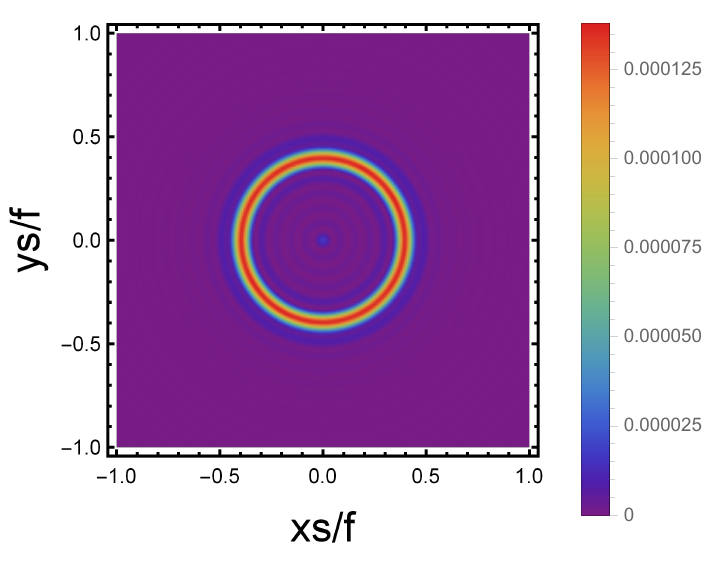}}
\subfigure[$u=0.5$]{
\includegraphics[scale=0.33]{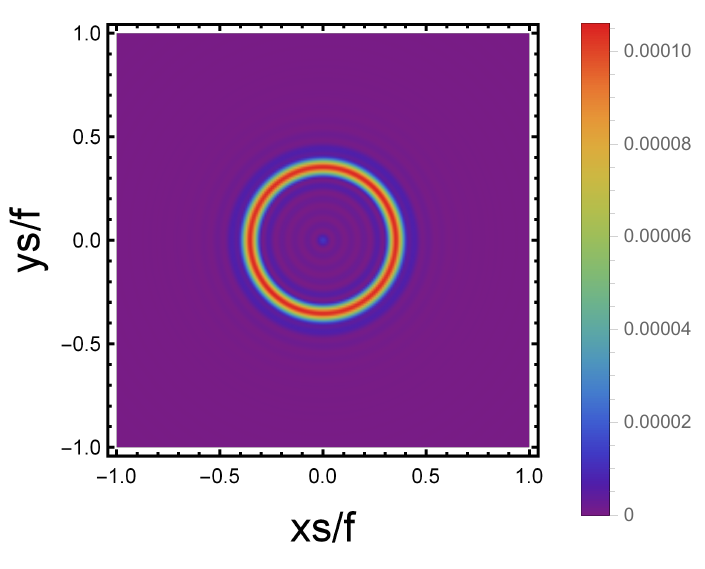}}
\subfigure[$u=0.7$]{
\includegraphics[scale=0.33]{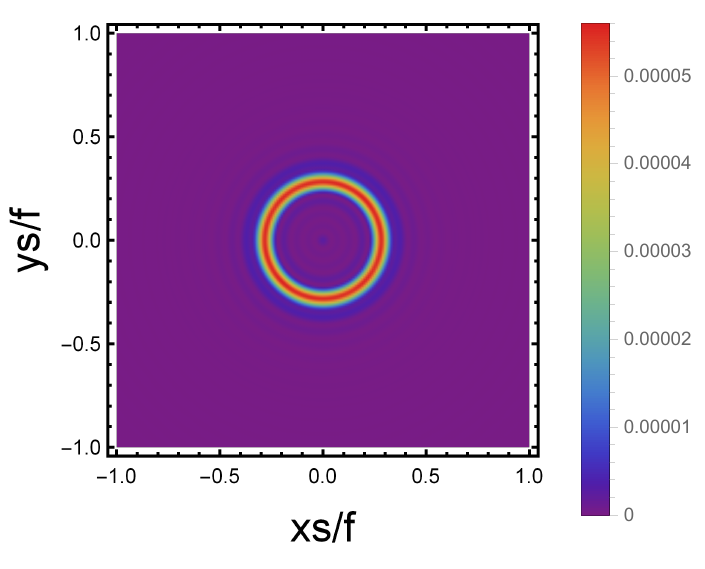}}
 \caption{The effect of chemical potential $u$ on the Einstein ring, when $\lambda=2$, $T=0.55$. } \label{fig77}
\end{figure}

Additionally, we explore the effect of frequency $\bar{\omega}$ on the Einstein ring under conditions when $Q=0.01$, $z_h=4$, $\lambda=2$ and $e=0.01$. Here, we find that higher $\bar{\omega}$ values mainly impact the shape of the ring, with larger $\bar{\omega}$ resulting in clearer ring, as illustrated in Fig. \ref{fig555}. That is,the lower the frequency, the greater the volatility. 
This observation aligns with the geometric optics approximation when $\bar{\omega}$ is large.

\begin{figure}[H]
\centering
\subfigure[$u=0.1$]{
\includegraphics[scale=0.45]{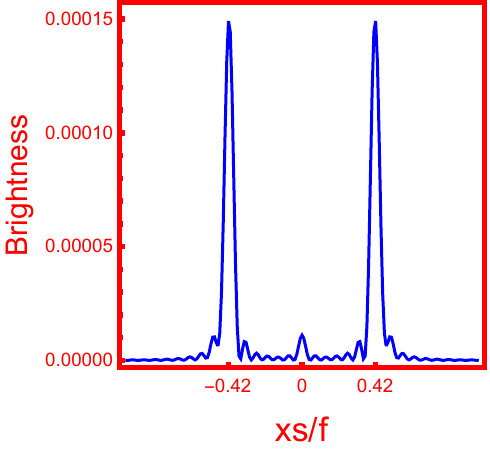}}
\subfigure[$u=0.3$]{
\includegraphics[scale=0.45]{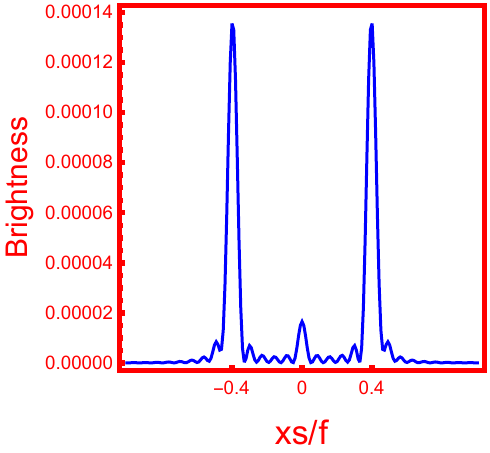}}
\subfigure[$u=0.5$]{
\includegraphics[scale=0.45]{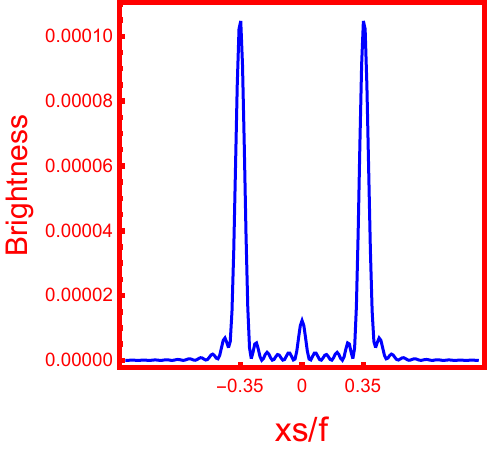}}
\subfigure[$u=0.7$]{
\includegraphics[scale=0.45]{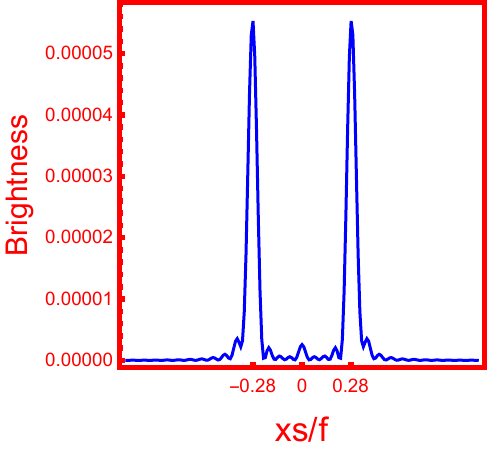}}
 \caption{The effect of chemical potential $u$ on the brightness, when $\lambda=2$, $T=0.55$.} \label{fig88}
\end{figure}

\begin{figure}[H]
\centering
\subfigure[$u=0.1$]{
\includegraphics[scale=0.33]{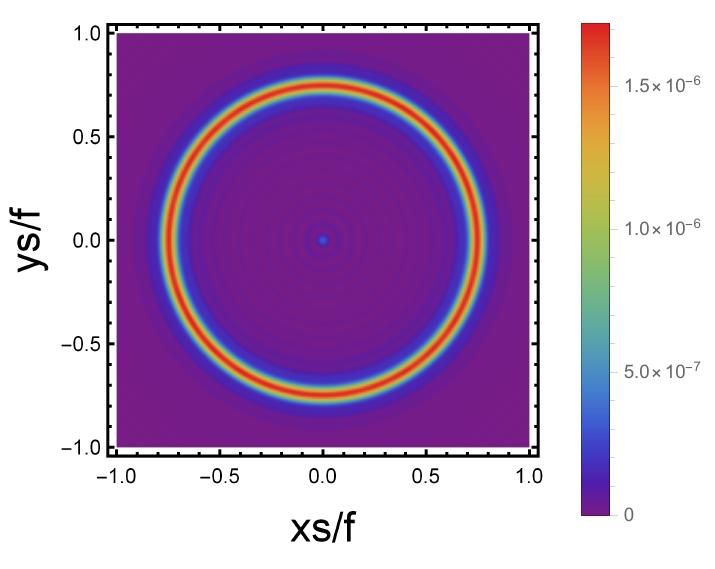}}
\subfigure[$u=0.2$]{
\includegraphics[scale=0.33]{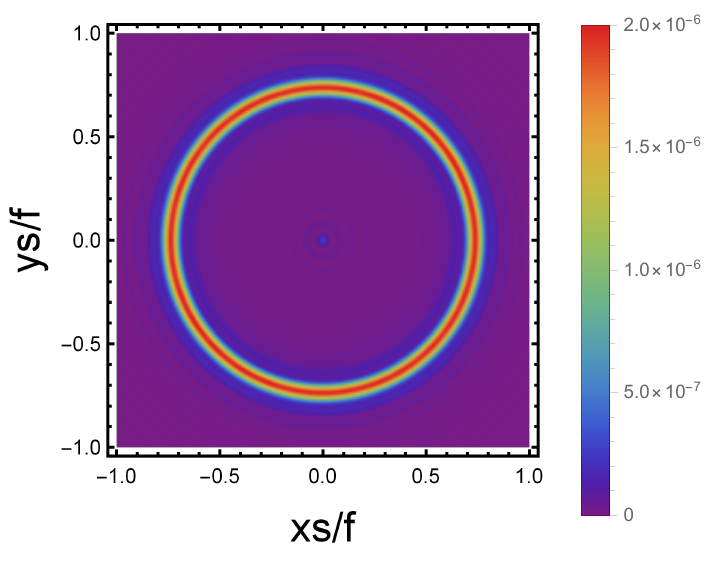}}
\subfigure[$u=0.3$]{
\includegraphics[scale=0.33]{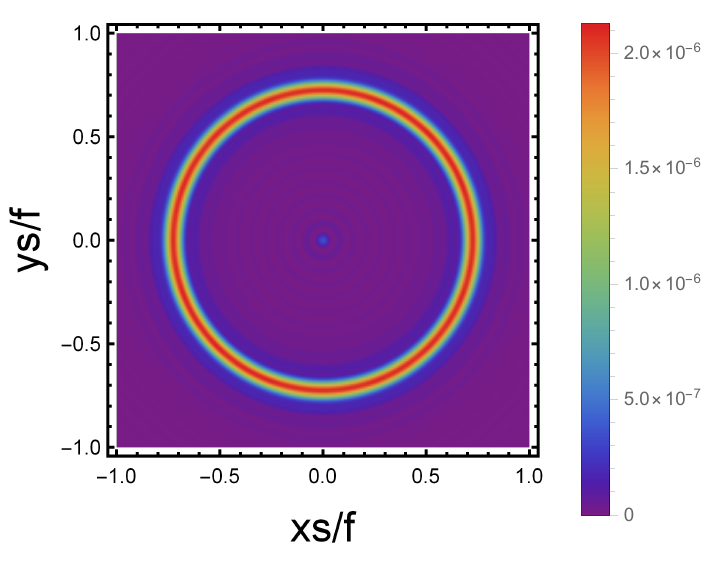}}
\subfigure[$u=0.4$]{
\includegraphics[scale=0.33]{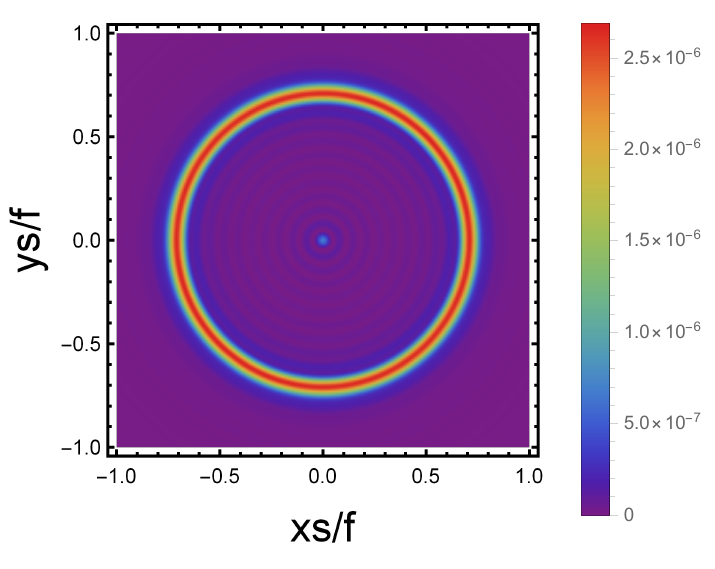}}
 \caption{The effect of chemical potential  $u$ on the Einstein ring, when $\lambda=4$, $T=0.3$.} \label{fig99}
\end{figure}
Further investigation into the effect of chemical potential $u$ on the Einstein ring for $\lambda=2$ and $T=0.55$ reveals that increasing $u$ leads to a reduction in the ring radius, as shown in Fig. \ref{fig77} and  Fig. \ref{fig88}. Especially in Fig. \ref{fig88}, we can see clearly that for $u=0.1, 0.3, 0.5, 0.7$, the Einstein ring locates at 0.42, 0.4, 0.35, 0.28, implying the increasing $u$ leads to a reduction in the ring radius. And we also find that the increasing $u$ also reduces the brightness of the Einstein ring. Explicitly, for $u=0.1,0.3,0.5,0.7$, the peaks of brightness are 0.00015, 0.00014, 0.00010, 0.00005 in order.

\begin{figure}[H]
\centering
\subfigure[$u=0.1$]{
\includegraphics[scale=0.45]{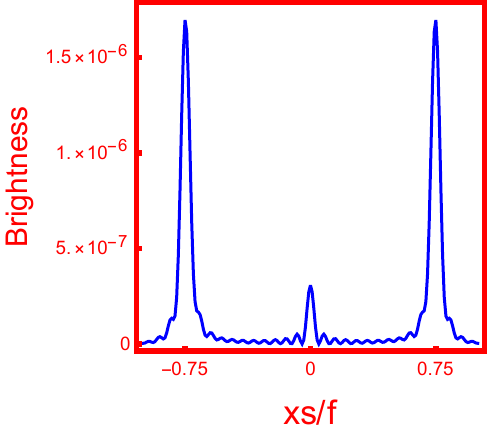}}
\subfigure[$u=0.2$]{
\includegraphics[scale=0.45]{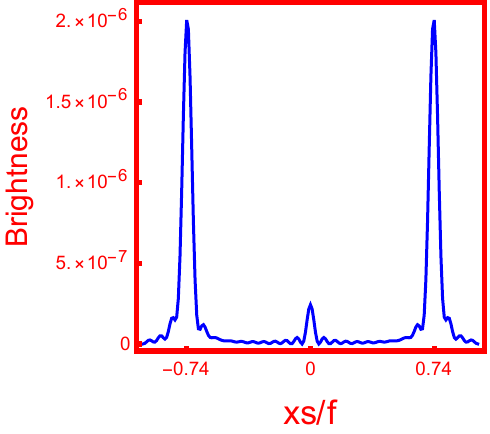}}
\subfigure[$u=0.3$]{
\includegraphics[scale=0.45]{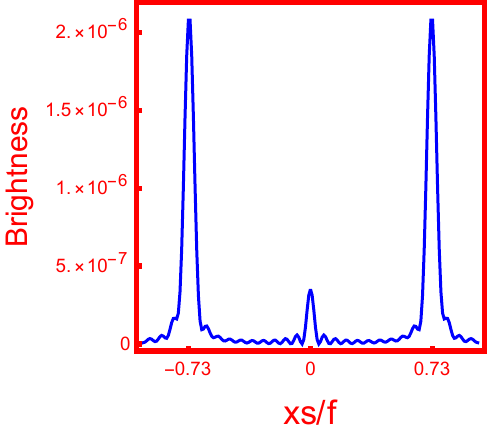}}
\subfigure[$u=0.4$]{
\includegraphics[scale=0.45]{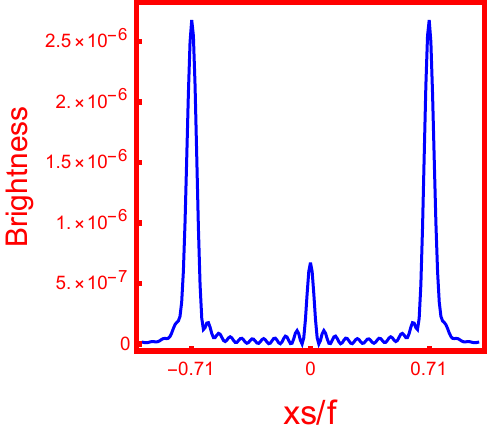}}
 \caption{The effect of chemical potential  $u$ on the brightness, when $\lambda=4$, $T=0.3$.} \label{fig100}
\end{figure}

\begin{figure}[H]
\centering
\subfigure[$T=0.00239$]{
\includegraphics[scale=0.33]{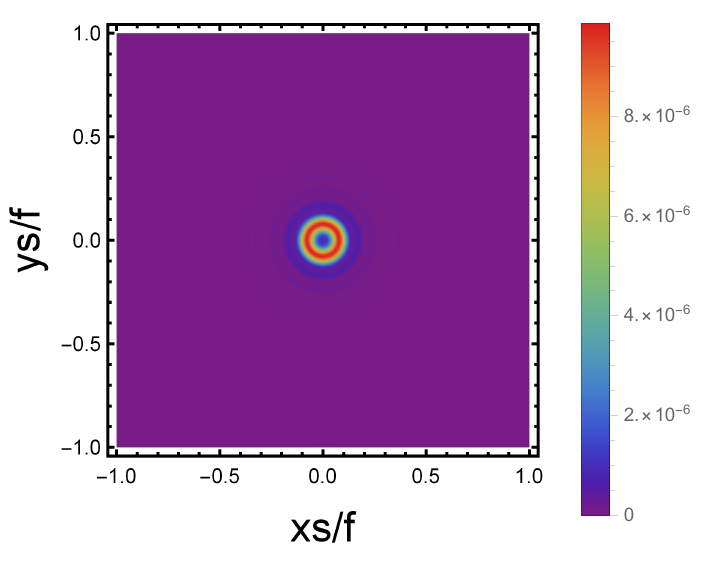}}
\subfigure[$T=0.01432$]{
\includegraphics[scale=0.33]{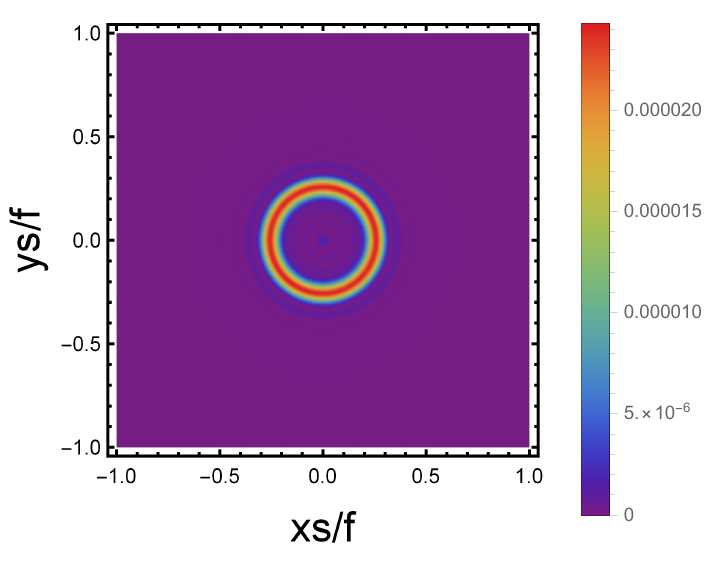}}
\subfigure[$T=0.02623$]{
\includegraphics[scale=0.33]{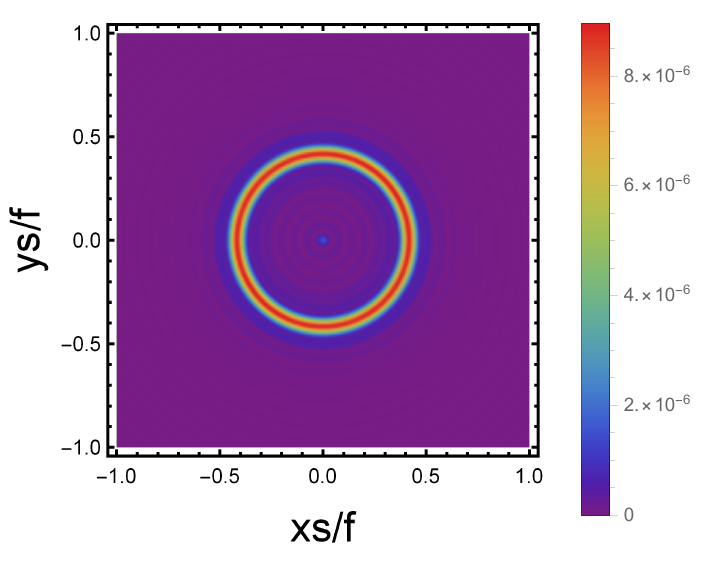}}
\subfigure[$T=0.03819$]{
\includegraphics[scale=0.33]{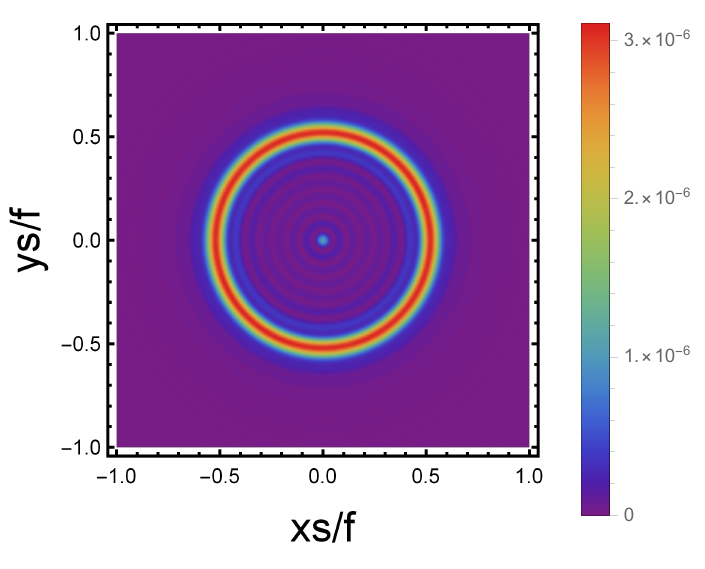}}
 \caption{The effect of  temperature $T$ on the Einstein ring,  when $\lambda=2$, $u=1$.} \label{fig11}
\end{figure}
Some results contrast with the previous studies under high temperature conditions, such as in \cite{Liu:2022cev}, where the ring radius remained unaffected with the changes in chemical potential due to the system approaching an RN black hole geometry. Here, at lower temperatures shown in  Fig. \ref{fig77} and  Fig. \ref{fig88}, respectively, we confirm that increasing $u$ consistently decreases the radius of the photon ring at lower temperature. For high temperatures, the results are the same with  \cite{Liu:2022cev}. For the high temperature case, the chemical potential does not affect the ring radius.

Similarly, for the case  $\lambda=4$, we also can plot the Einstein ring and the brightness, shown in  Fig. \ref{fig99} and 
Fig. \ref{fig100}.  From Fig. \ref{fig100}, we can see  that for $u=0.1, 0.2, 0.3, 0.4$, the Einstein ring locates at  0.75, 0.74, 0.73, 0.71. The pattern of change is exactly the same
as the case $\lambda=2$. But it is worth pointing out that the change of photon ring brightness for $\lambda=4$ is different from the case of $\lambda=2$. In the case $\lambda=4$, as the chemical potential increases from u=0.1 to u=0.4, the brightness of the photon ring also increases, not decreases.

\begin{figure}[H]
\centering
\subfigure[$T=0.00239$]{
\includegraphics[scale=0.45]{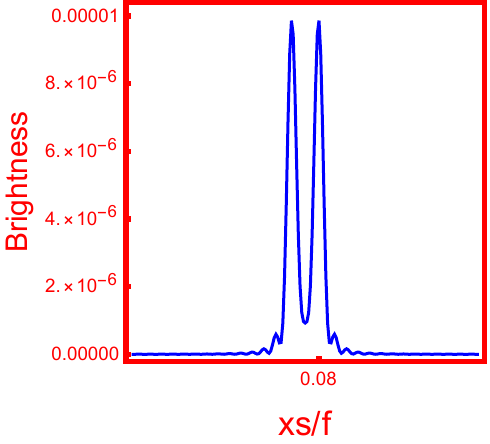}}
\subfigure[$T=0.01432$]{
\includegraphics[scale=0.45]{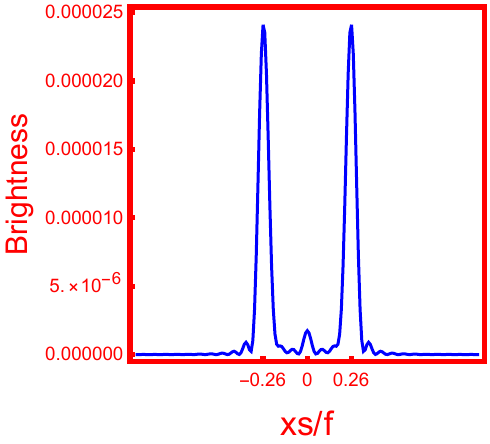}}
\subfigure[$T=0.02623$]{
\includegraphics[scale=0.45]{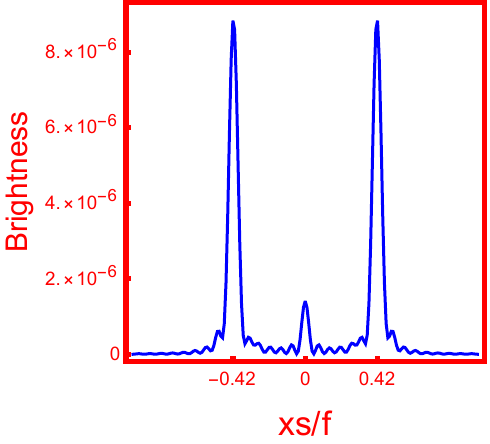}}
\subfigure[$T=0.03819$]{
\includegraphics[scale=0.45]{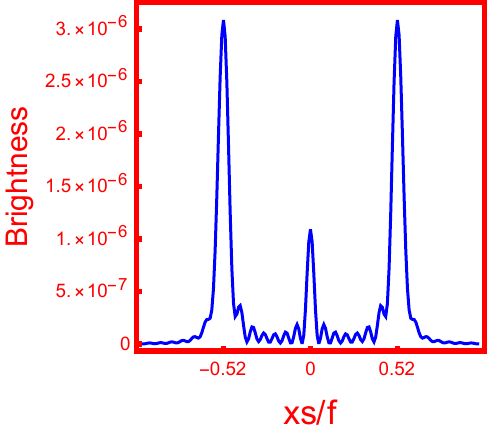}}
 \caption{The effect of  temperature $T$ on the brightness, when $\lambda=2$, $u=1$.} \label{fig12}
\end{figure}

\begin{figure}[H]
\centering
\subfigure[$T=0.81169$]{
\includegraphics[scale=0.33]{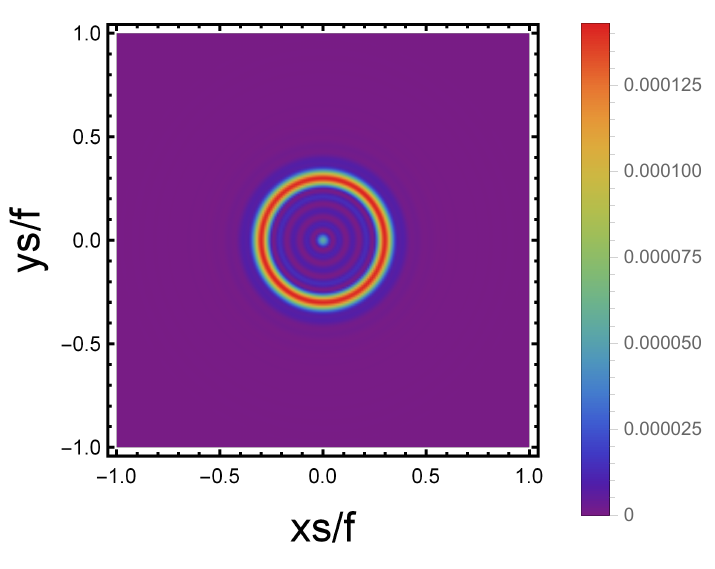}}
\subfigure[$T=0.44165$]{
\includegraphics[scale=0.33]{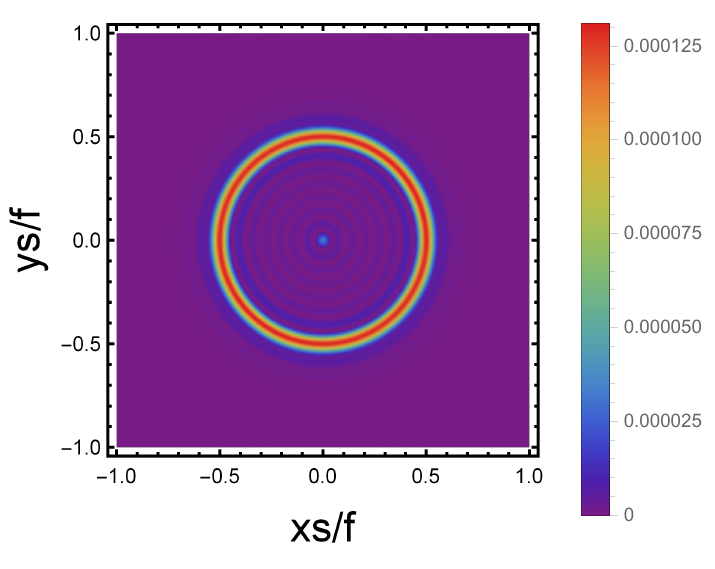}}
\subfigure[$T=0.33423$]{
\includegraphics[scale=0.33]{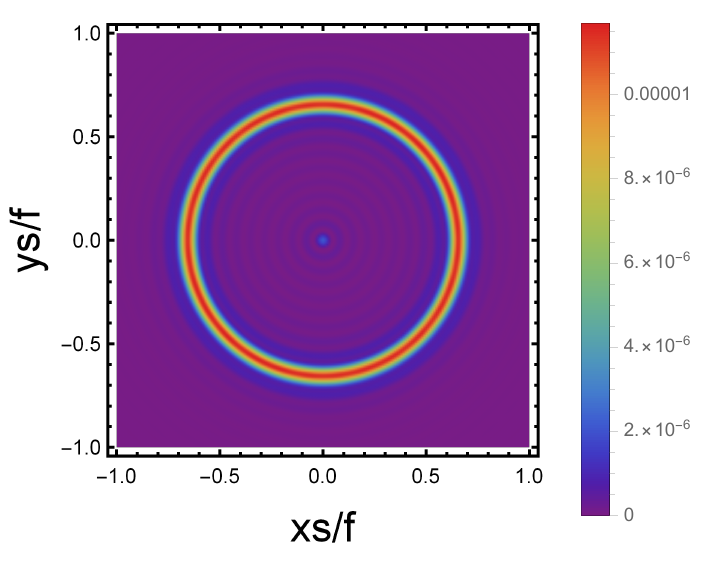}}
\subfigure[$T=0.29245$]{
\includegraphics[scale=0.33]{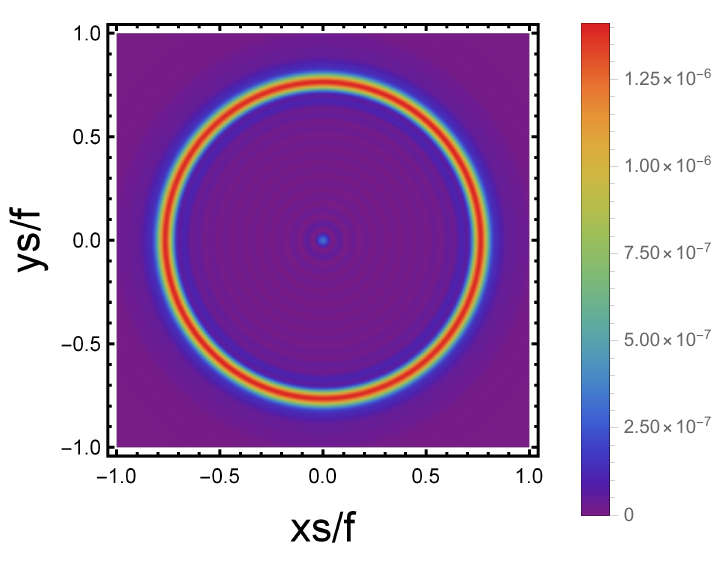}}
 \caption{The effect of  temperature $T$ on the Einstein ring  when   $\lambda=2$, $u=0.1$. } \label{fig133}
\end{figure}

\begin{figure}[H]
\centering
\subfigure[$T=0.81169$]{
\includegraphics[scale=0.45]{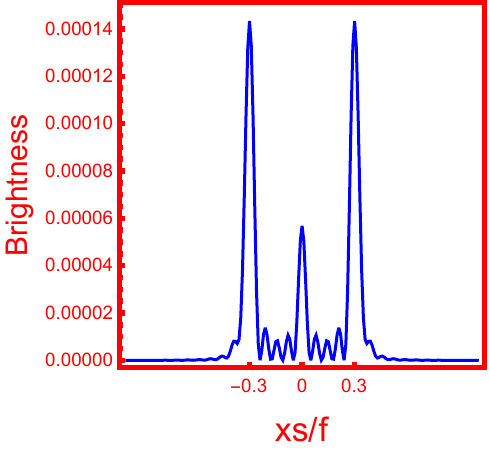}}
\subfigure[$T=0.44165$]{
\includegraphics[scale=0.45]{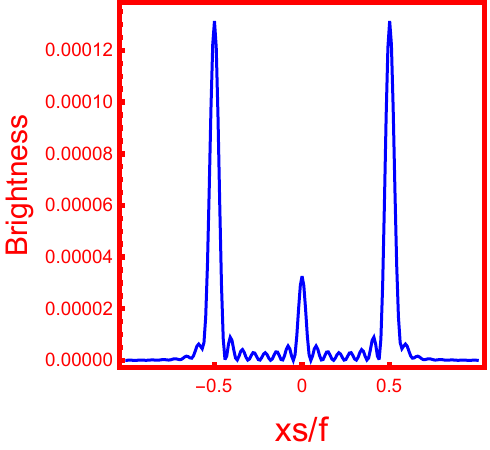}}
\subfigure[$T=0.33423$]{
\includegraphics[scale=0.45]{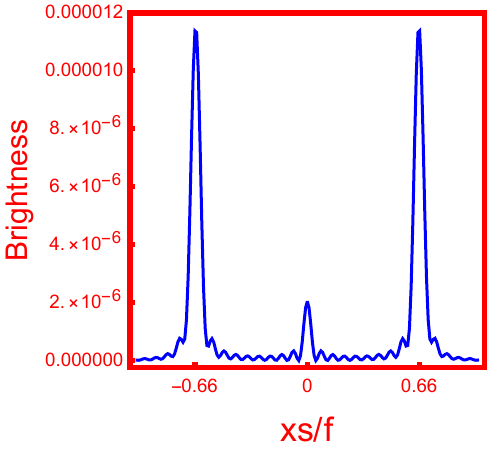}}
\subfigure[$T=0.29245$]{
\includegraphics[scale=0.45]{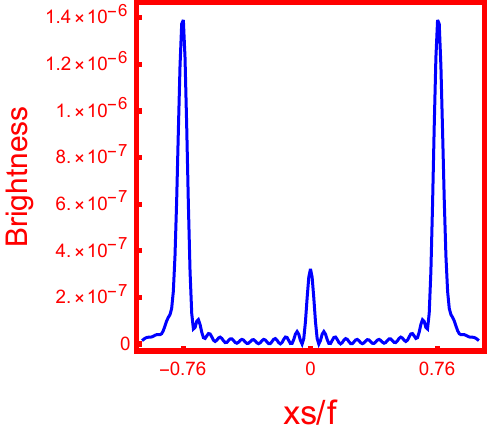}}
 \caption{The effect of  temperature $T$ on the brightness, where $\lambda=2$, $u=0.1$.} \label{fig14}
\end{figure}

\begin{figure}[H]
\centering
\subfigure[$T=0.79578$]{
\includegraphics[scale=0.3]{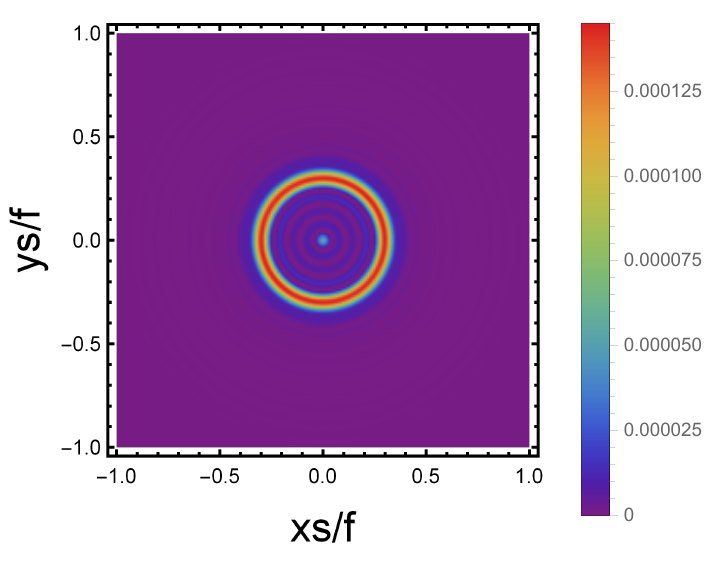}}
\subfigure[$T=0.50131$]{
\includegraphics[scale=0.3]{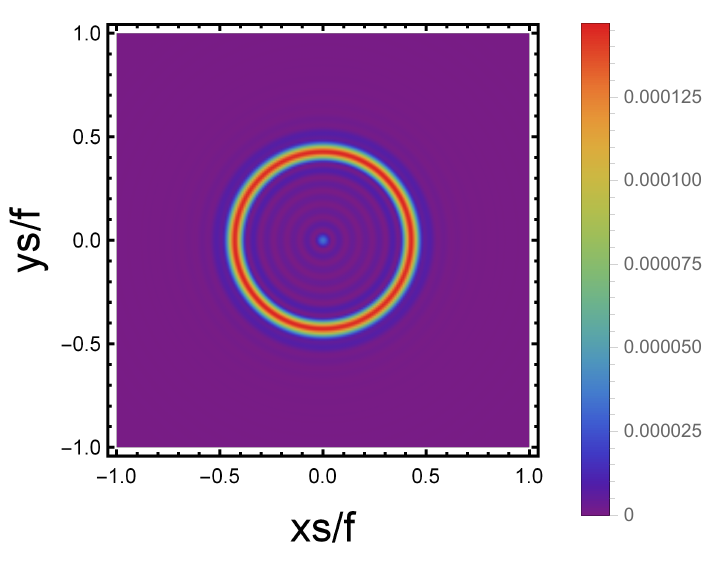}}
\subfigure[$T=0.41560$]{
\includegraphics[scale=0.3]{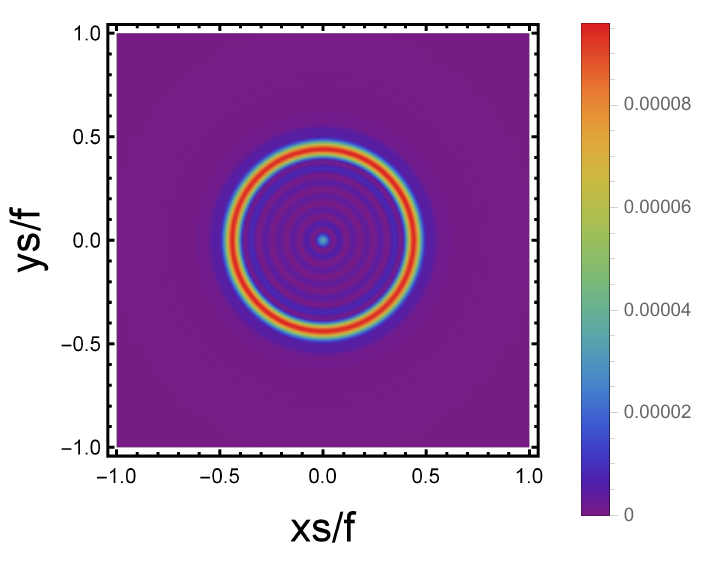}}
\subfigure[$T=0.36937$]{
\includegraphics[scale=0.3]{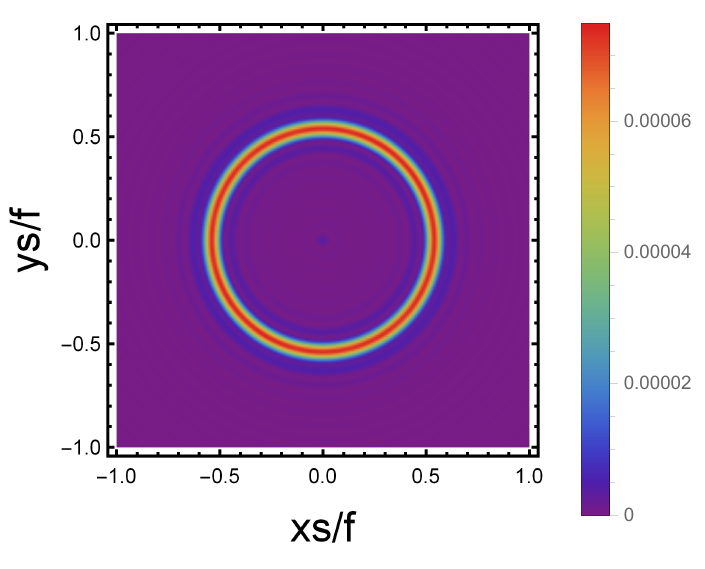}}
 \caption{The effect of temperature on the Einstein ring when  $\lambda=4$, $u=1$.} \label{fig15}
\end{figure}

\begin{figure}[H]
\centering
\subfigure[$T=0.79578$]{
\includegraphics[scale=0.45]{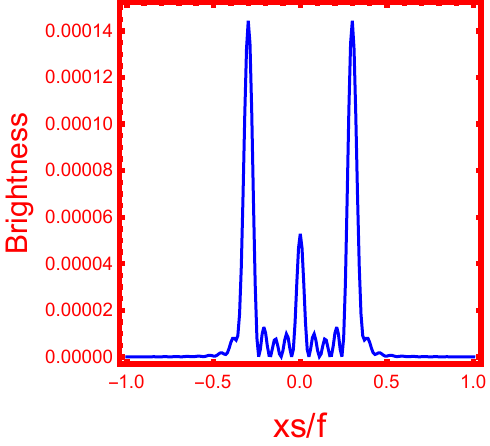}}
\subfigure[$T=0.50131$]{
\includegraphics[scale=0.45]{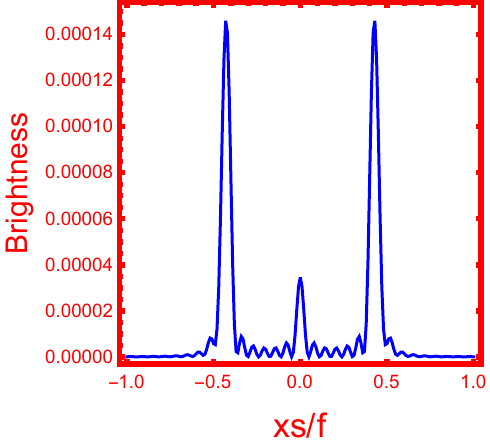}}
\subfigure[$T=0.41560$]{
\includegraphics[scale=0.45]{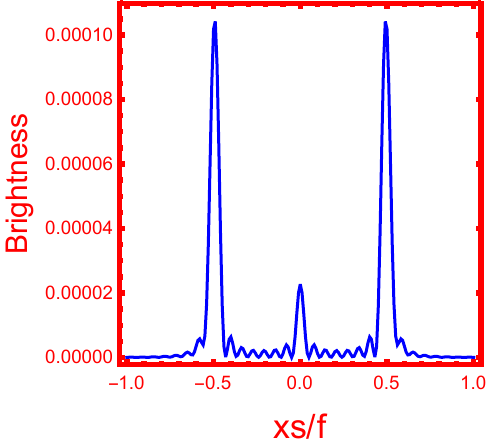}}
\subfigure[$T=0.36937$]{
\includegraphics[scale=0.45]{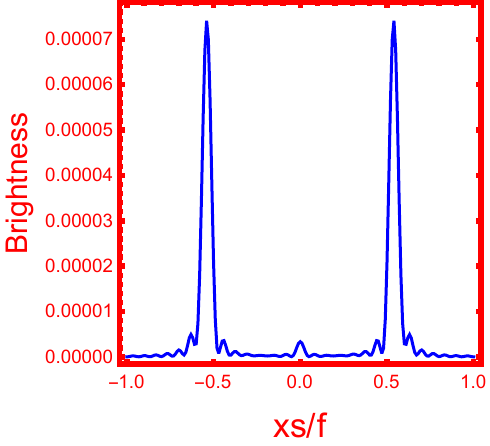}}
 \caption{The effect of temperature on the brightness when  $\lambda=4$, $u=1$.} \label{fig16}
\end{figure}

\begin{figure}[H]
\centering
\subfigure[$T=0.357968$]{
\includegraphics[scale=0.29]{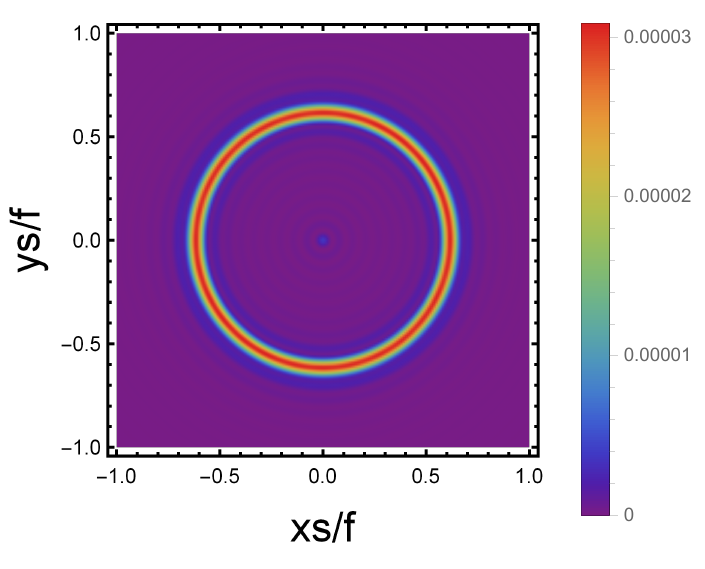}}
\subfigure[$T=0.303956$]{
\includegraphics[scale=0.29]{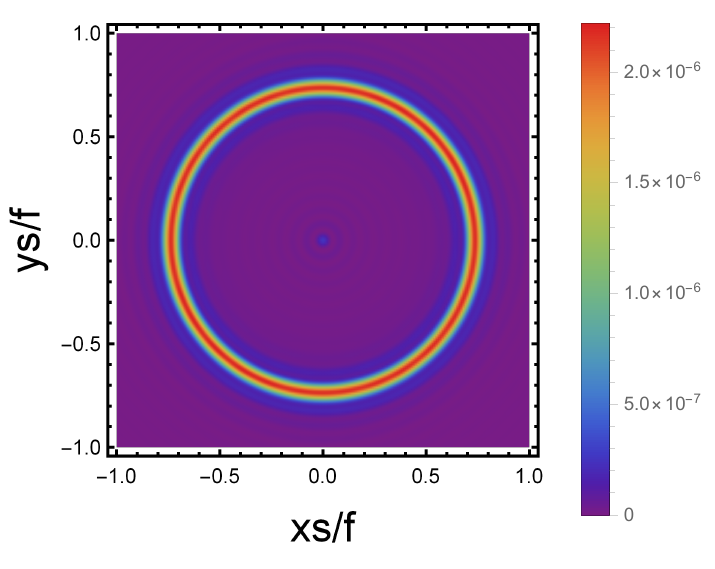}}
\subfigure[$T=0.28755$]{
\includegraphics[scale=0.29]{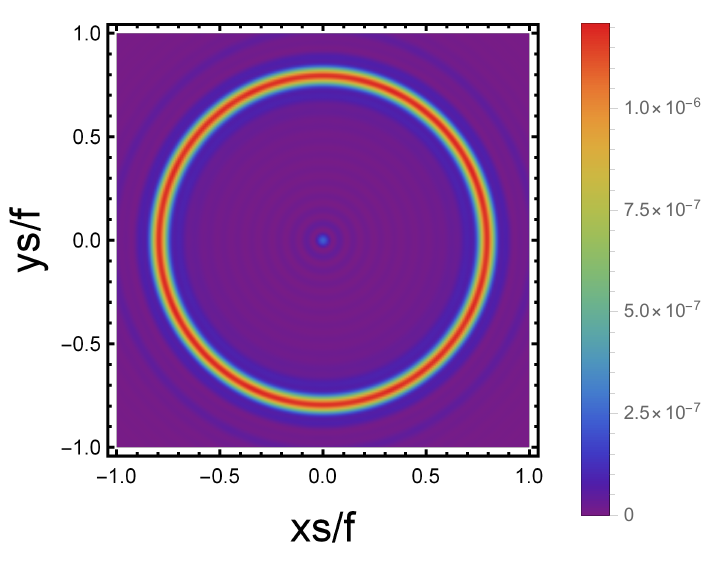}}
\subfigure[$T=0.28047$]{
\includegraphics[scale=0.29]{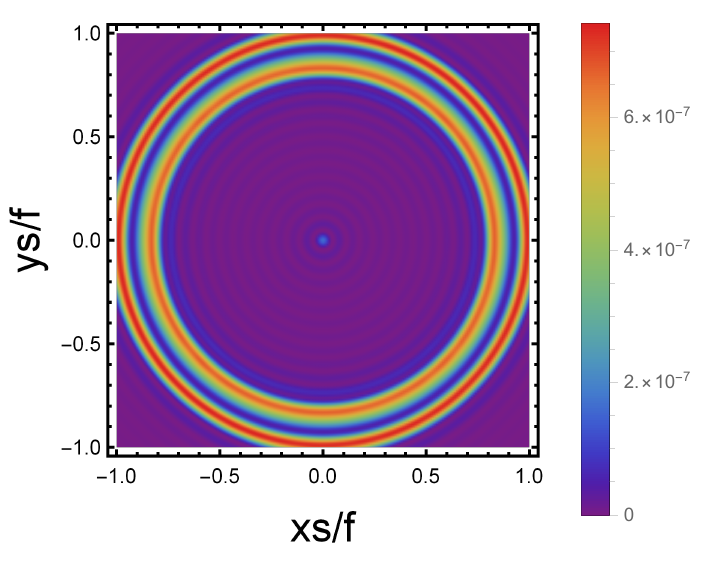}}
 \caption{The effect of temperature on the Einstein ring when  $\lambda=4$, $u=0.1$.} \label{fig1500}
\end{figure}

\begin{figure}[H]
\centering
\subfigure[$T=0.357968$]{
\includegraphics[scale=0.45]{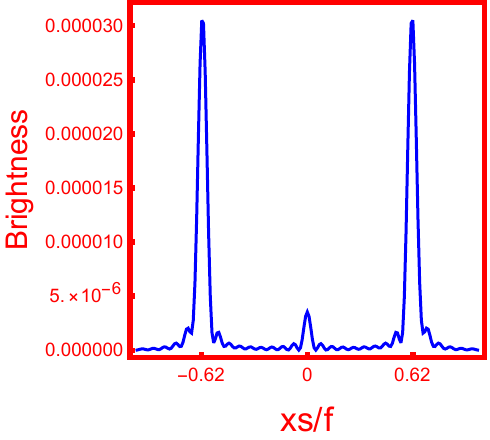}}
\subfigure[$T=0.303956$]{
\includegraphics[scale=0.45]{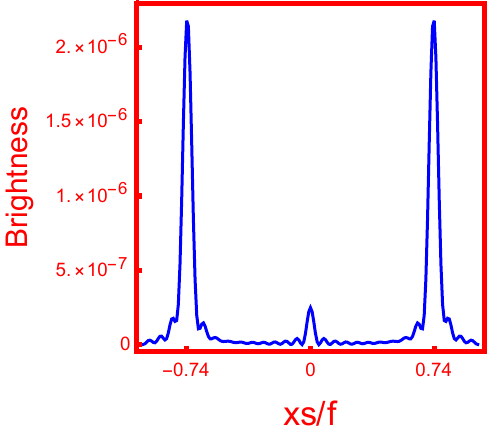}}
\subfigure[$T=0.28755$]{
\includegraphics[scale=0.45]{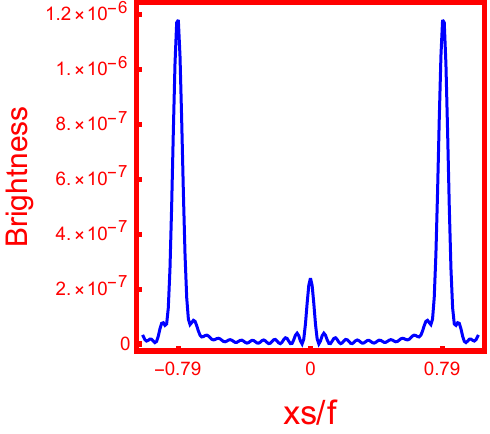}}
\subfigure[$T=0.28047$]{
\includegraphics[scale=0.45]{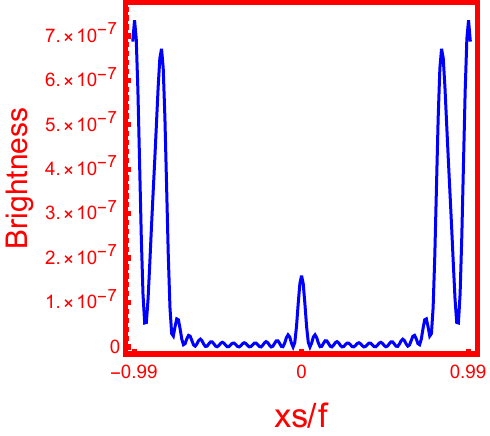}}
 \caption{The effect of temperature on the brightness when  $\lambda=4$, $u=0.1$.} \label{fig160}
\end{figure}

From the previous analysis, we observe that the chemical potential $u$ has a consistent effect on the radius of the ring across different values of $\lambda$. Next, we explore the influence of temperature $T$ on the radius of the Einstein ring for various $\lambda$, as illustrated in Fig. \ref{fig11} to Fig. \ref{fig160}.


Specifically, in Fig. \ref{fig11} and Fig. \ref{fig133}, we examine the effect of temperature $T$ on the Einstein ring with $\lambda=2$, considering different chemical potentials $u=1$ (Fig. \ref{fig11}, Fig. \ref{fig12}) and $u=0.1$ (Fig. \ref{fig133}, Fig. \ref{fig14}). We find that for $\lambda=2$ and $u=1$, as temperature $T$ increases, the ring radius also increases. From Fig. \ref{fig12}, we clearly see for $T=0.00239, 0.01432, 0.02623, 0.03819$, the Einstein ring locates at 0.08, 0.26, 0.42 and 0.52. Conversely, for $u=0.1$, the ring radius decreases with increasing temperature, as depicted in Fig. \ref{fig133} and Fig. \ref{fig14}. Clearly for $T=0.81169, 0.44165, 0.33423, 0.29245$, the Einstein ring locates at 0.3, 0.5, 0.66, 0.76 in Fig. \ref{fig14}.
In other words, how does the temperature affect the location of the ring radius depending on the value of the chemical potential for the charged black hole, which has not been mentioned in previous literature.

Similarly, in Fig. \ref{fig15} and Fig. \ref{fig1500}, we analyze the effect of temperature $T$ on the Einstein ring for $\lambda=4$, considering chemical potentials $u=1$ (Fig. \ref{fig15},  Fig. \ref{fig16}) and $u=0.1$ (Fig. \ref{fig1500}, 
 Fig. \ref{fig160}). Here, we observe that for both $u=1$ and $u=0.1$, the ring radius decreases as temperature $T$ increases across Fig. \ref{fig16} and Fig. \ref{fig160}. This conclusion is completely different from the case of RN-AdS black hole ($\lambda=2$), where the effect of temperature on the radius of the black hole ring does not depend on the chemical potential of the black hole. {The reason is that the  temperature in Eq. (\ref{tr})
 includes $\lambda$ order terms of event horizon $r_h$. For different $\lambda$ , the dominant terms differences. So the temperature dependence behavior of the photon ring radius may significantly change for different parameter $\lambda$.}

In summary, our investigation reveals distinct temperature dependencies on the ring radius for different combinations of $\lambda$ and $u$. Specifically, for $\lambda=2$ and $u=1$, increasing temperature $T$ leads to an increase in the ring radius, whereas for $\lambda=2$ and $u=0.1$, the ring radius decreases with rising temperature. Conversely, for $\lambda=4$, the temperature increase consistently results in a decrease in the ring radius, regardless of the chemical potential $u$. These findings highlight the nuanced interplay between the temperature and gravitational parameters in shaping the characteristics of Einstein rings in the context of charged black holes in Lorentz symmetry breaking massive gravity.

\section{Geometric optics }
\label{3}
To understand the images of the black holes more intuitively, we will study photon trajectories by the following Euler-Lagrange equation
\begin{eqnarray}
\frac{d}{d\beta 
}\left(\frac{\partial \mathcal{L}}{\partial \dot{x}^{\mu}}\right)=\frac{\partial \mathcal{L}}{\partial x^{\mu}},
\label{eleq}
\end{eqnarray}
in which $\beta$ is the affine parameter, $\dot{x}^{\mu}$ is the four-velocity of the light ray and $\mathcal{L}$ is the Lagrangian, taking the form as
\begin{eqnarray}
\mathcal{L}&=&\frac{1}{2}g_{\mu\nu}(\dot{x}^{\mu}-e g^{\mu \nu }A_{\nu})(\dot{x}^{\nu}-e g^{\mu \nu }A_{\mu}) \nonumber \\
 &=&\frac{1}{2}\left( - G(r)(\dot{t}+e \frac{A_t}{G(r)})^2+\frac{\dot{r}^2}{G(r)}+r^2\left(\dot{\psi}^2+\sin^2{\psi}~ \dot{\phi}^2 \right)\right).\nonumber\\
\label{laeq}
\end{eqnarray}
We will consider the case  $\psi = \pi/2$, which indicates that the light ray always moves on the equatorial plane. In addition, due to the spacetime is spherically symmetric, there are two conserved quantities, namely the energy $E$,  and angular momentum $L$,
which can be defined as 
\begin{equation}
\bar{\omega}=-\frac{\partial \mathcal{L}}{\partial \dot{t}}=G(r) \dot{t}+e A_t,
 \label{laeqe}
\end{equation}   
\begin{equation}
L=\frac{\partial \mathcal{L}}{\partial \dot{\phi}}=r^2 \dot{\phi}.\label{laeql}
\end{equation}  
  
Combining Eqs.(\ref{metric}), (\ref{laeq}), (\ref{laeqe})  and  (\ref{laeql}), the time, azimuthal and radial component of the four-velocity can be expressed as
\begin{eqnarray}
  &&\dot{t}=\frac{\bar{\omega}-e A_t}{G(r)},  \label{time} \\
  &&\dot{\phi}= \frac{L}{r^2}, \label{psi} \\
  &&\dot{r}^2+\frac{G(r) L^2}{r^2}-\bar{\omega}^2=0.
 \label{radial}
\end{eqnarray}
Note that here we have used the condition $ds^2=0$.   Eq.(\ref{radial}) can further   be rewritten as
\begin{equation}
 \dot{r}^2=\bar{\omega}^2-L^2 V(r),  \label{vbr}
\end{equation}
where
\begin{equation}
V(r)=\frac{1}{r^2} (1-\frac{2M}{r}-\chi\frac{Q^2}{r^{\lambda}}+\frac{ r^2}{l^2}),
\end{equation} \label{epotential}
is an effective potential.
At the  photon sphere, the motion equations of the light ray   satisfy  $\dot{r}=0$, and  $\ddot{r}=0$, which also means 
\begin{eqnarray}
V(r)=\frac{\omega}{L^2},  V^{'}(r) = 0,
\label{condition1}
\end{eqnarray}
where the  prime denotes the first derivative with respect to the radial coordinate $r$. Based on Eq.(\ref{condition1}), we can obtain the maximum of the effective potential and the 
corresponding radius coordinate. For the case $\lambda =-1,\lambda =2$, we find 
\begin{eqnarray}
V_{max}&=&1 +\frac{256 Q^2 r_h^4}{\Pi^4}-\frac{64r_h^2 \left(Q^2+r_h^4+r_h^2\right)}{\Pi^3}+\frac{ 16r_h^2}{\Pi^2}, \nonumber \\r_{max}&=&\frac{ \Pi}{4r_h},
\end{eqnarray}
in which 
 \begin{eqnarray}
\Pi=\sqrt{\left(3 Q^2+3 r_h^4+3 r_h^2\right)^2-32 Q^2 r_h^2}+3 Q^2+3 r_h^4+3r_h^2. \nonumber\\
\end{eqnarray}
For the case $Q=0$, the result is consistent with~\cite{Ovgun:2023ego}. 

\begin{figure}[H]
\centering
\subfigure[$\lambda=2$]{
\includegraphics[scale=0.35]{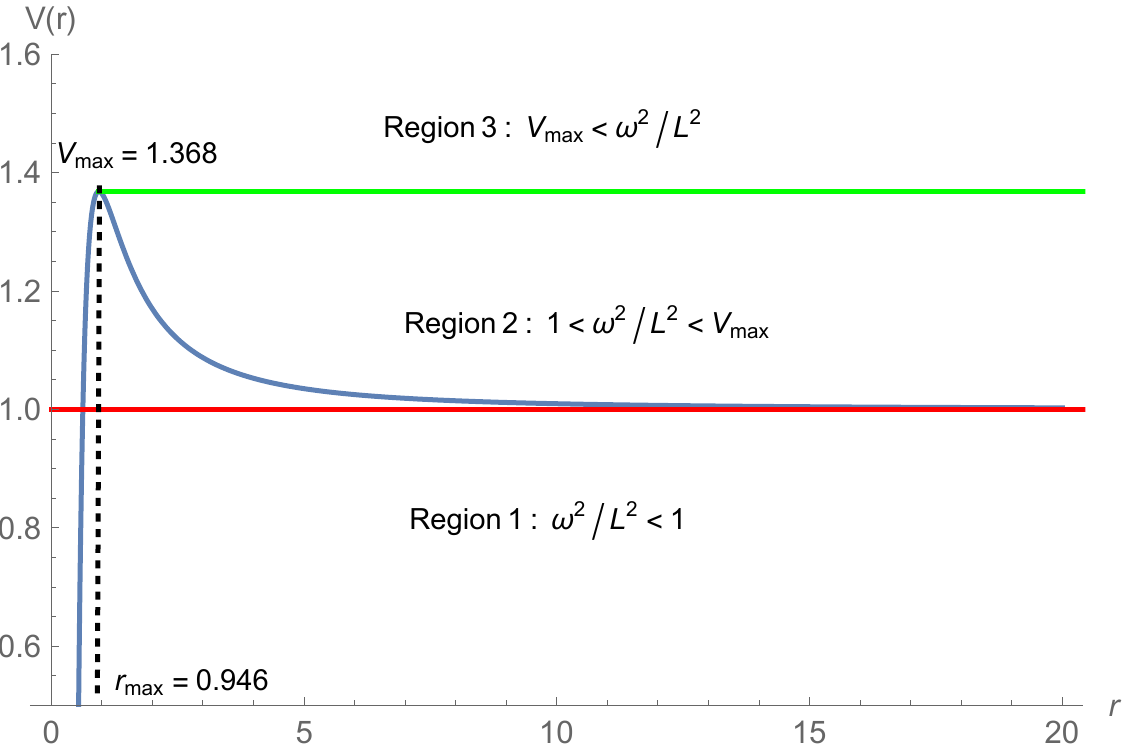}}
\subfigure[$\lambda=4$]{
\includegraphics[scale=0.35]{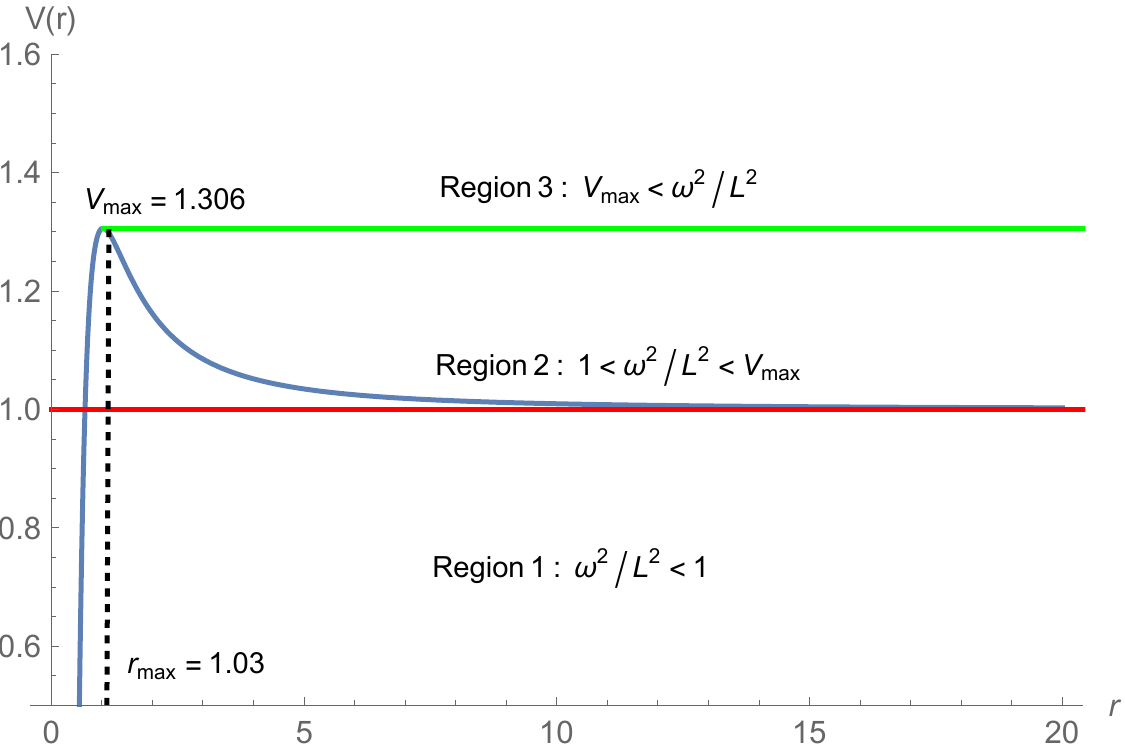}}
 \caption{The profile of the effective potential for different $\lambda$ with   $r_h=0.5$, $\lambda =-1$ and $Q=0.1$.} \label{fig25}
\end{figure}

In Figure \ref{fig25}, the effective potential is plotted for various $\lambda$ values under fixed parameters. Notably, it vanishes at the event horizon and rises sharply with the radial distance. It peaks at the photon sphere, subsequently declining stepwise until stabilizing at a fixed value. This potential profoundly influences the trajectory of light rays moving radially inward. Region 1 is inaccessible because angular momentum cannot exceed total energy. Photons initiating motion with $1 < \bar{\omega}^2/L^2 < V_{\text{max}}$ encounter a potential barrier (region 2) and are reflected outward. If $\bar{\omega}^2/L^2 > V_{\text{max}}$, photons fall directly into the black hole without encountering this barrier. At $\bar{\omega}^2/L^2 = V_{\text{max}}$, light orbits the black hole indefinitely due to non-zero angular velocity, forming a circular and unstable orbit.

 This study focuses on region 2, where photons enter with an ingoing angle $\psi_{\text{in}}$, defined as
  ~\cite{
Hashimoto:2018okj,Hashimoto:2019jmw}
\begin{eqnarray}
 \label{lac}
\cos\psi_{in}=\frac{g_{ij}\bar{u}^i \bar{n}^j}{|\bar{u}||\bar{n}|}|_{r=\infty},
\end{eqnarray}
where $g_{ij}$ is the induced metric on a constant time slice,  $\bar{u}^i$ is the spatial component of the four-velocity,  $\bar{n}^j$ is the normal vector to the boundary, and  $|\bar{u}|,|\bar{n}|$ are the norms of  $\bar{u}^i$, $\bar{n}^j$ respectively. 

\begin{figure}[H]
	\centering
	\includegraphics[trim=2.2cm 2.9cm 2.4cm 1.1cm, clip=true, scale=0.8]{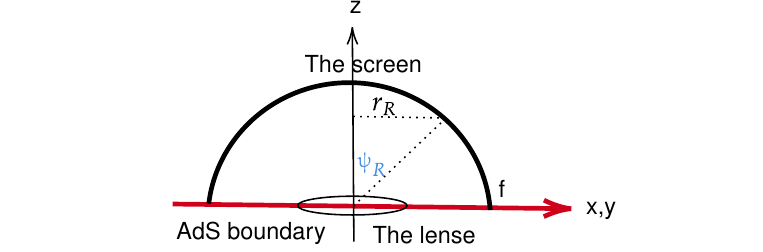}
	\caption{The  geometric relation   between the ring angle  $\psi_R $ and ring radius  $r_R$.}\label{inout}
\end{figure}

With Eqs.(\ref{metric}), (\ref{time}), (\ref{psi}),  (\ref{radial}) and (\ref{lac}), after calculation, we can explicitly obtain 
\begin{eqnarray}
\sin\psi_{in}=\frac{L}{\bar{\omega}}.
\end{eqnarray} \label{xyz}
This relation is also valid when the  photon is located at the photon ring, where  the angular momentum ls labeled as $L_p$.  In this case, we know 
\begin{eqnarray}
\sin\psi_{in}=\frac{L_p}{\bar{\omega}}.
\end{eqnarray} \label{lpp}

In addition, from Fig.~\ref{inout}, we find the angle of the ring can be expressed by the following relation~\cite{Hashimoto:2018okj,Hashimoto:2019jmw}
\begin{eqnarray}
\text{}\sin\psi_{R}=\frac{{r_R}}{ f}. 
\end{eqnarray}
Physically, $\psi_{in}$ and $\psi_{R}$ correspond to the angle of the photon ring, and they in principle should be the same. We will verify this conclusion by numerical method below.

\begin{figure}[H]
\centering
\subfigure[$u=0.1$]{
\includegraphics[scale=0.42]{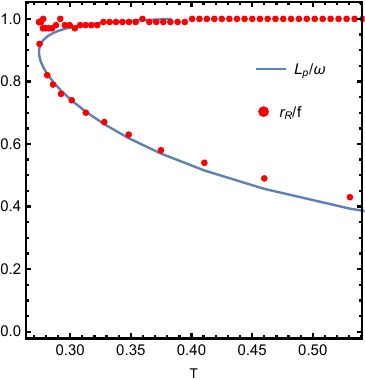}}
\subfigure[$u=0.5$]{
\includegraphics[scale=0.42]{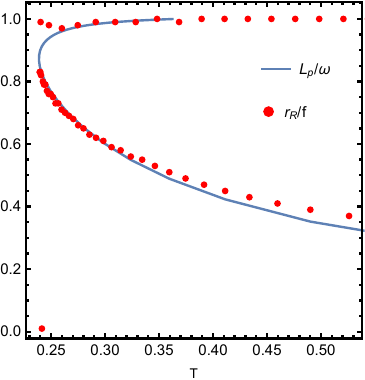}}
\subfigure[$u=1$]{
\includegraphics[scale=0.42]{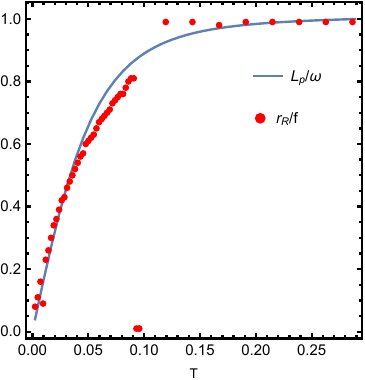}} \caption{Comparison of the results obtained by geometric optics and wave optics at different chemical potentials when $\lambda=2$.} \label{fig17}
\end{figure}

\begin{figure}[H]
\centering
\subfigure[$T=0.35$]{
\includegraphics[scale=0.42]{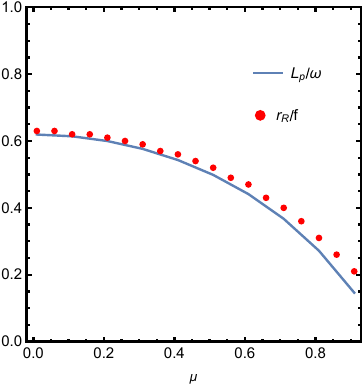}}
\subfigure[$T=0.55$]{
\includegraphics[scale=0.42]{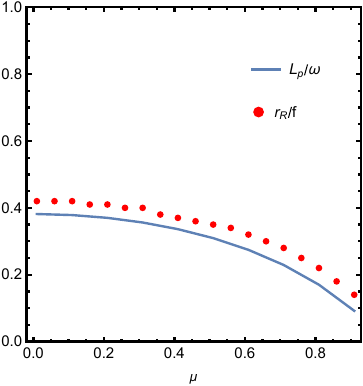}}
\subfigure[$T=0.95$]{
\includegraphics[scale=0.42]{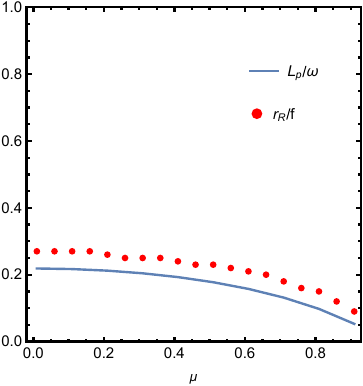}}
\caption{Comparison of the results obtained by geometric optics and wave optics at different temperature when $\lambda=2$.} \label{fig18}
\end{figure}

\begin{figure}[H]
\centering
\subfigure[$u=0.1$]{
\includegraphics[scale=0.32]{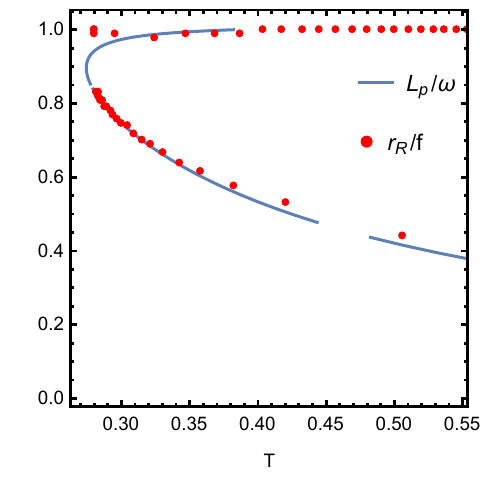}}
\subfigure[$u=0.5$]{
\includegraphics[scale=0.32]{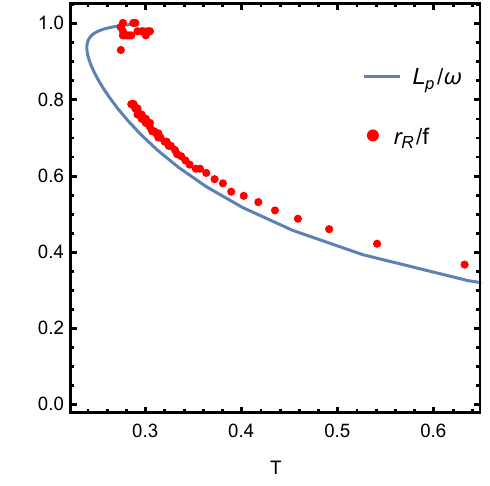}}
\subfigure[$u=1$]{
\includegraphics[scale=0.32]{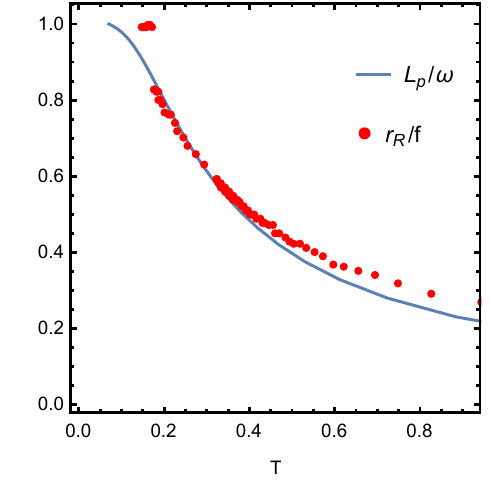}}
 \caption{Comparison of the results obtained by geometric optics and wave optics at different chemical potentials when $\lambda=4$.} \label{fig19}
\end{figure}

\begin{figure}[H]
\centering
\subfigure[$T=0.30$]{
\includegraphics[scale=0.32]{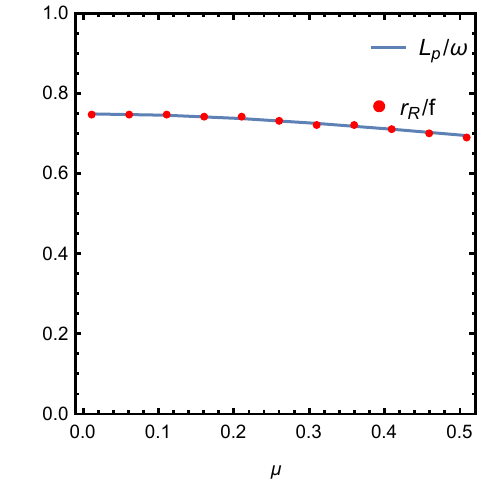}}
\subfigure[$T=0.55$]{
\includegraphics[scale=0.32]{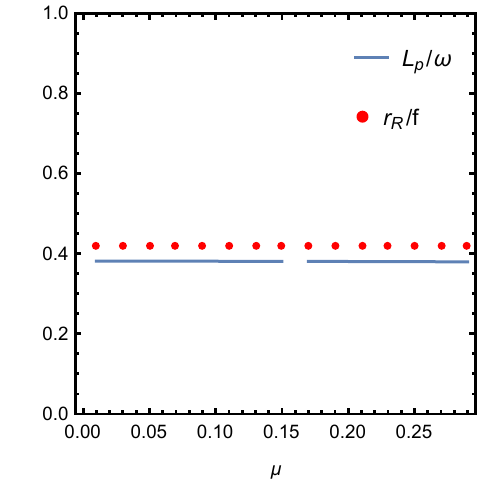}}
\subfigure[$T=0.95$]{
\includegraphics[scale=0.32]{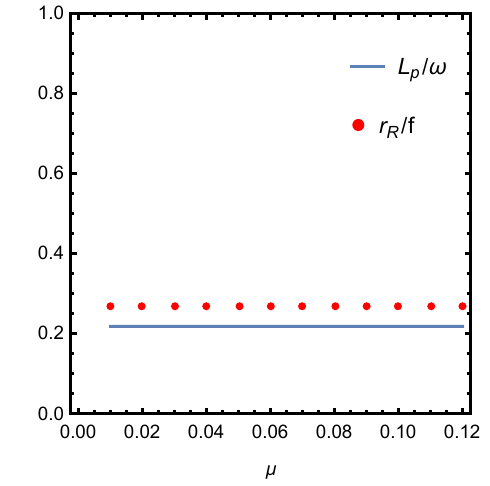}} \caption{Comparison of the results obtained by geometric optics and wave optics at different temperature when $\lambda=4$.} \label{fig20}
\end{figure}

In Fig.~\ref{fig17}, we investigate the variation of the Einstein ring radius $r_R$ with temperature $T$ for different chemical potentials $u=0.1$, $u=0.5$, and $u=1$, shown in subfigures (a), (b), and (c) respectively, when $\lambda=2$. The radius of the Einstein ring $r_R$ in units of $f$ is depicted by discrete red dots, while the solid curves represent results derived from geometric optics. We observe that when the chemical potential is small, such as $u=0.1$ and $u=0.5$, the ring radius does not exhibit a monotonic relationship with increasing temperature $T$. However, as the chemical potential $u$ increases to $u=1$, the ring radius monotonically increases with temperature.

In Fig.~\ref{fig18}, we examine the changes in the Einstein ring radius $r_R$ with different chemical potentials $u$ for fixed temperatures $T=0.35$, $T=0.55$, and $T=0.95$, presented in subfigures (a), (b), and (c) respectively, when $\lambda=2$. Similar to Fig.~\ref{fig17}, the radius of the Einstein ring $r_R$ is shown with discrete red dots, while the solid curves denote geometric optics predictions. From Fig.~\ref{fig18}, it is evident that regardless of the temperature $T$, the radius of the Einstein ring decreases as the chemical potential $u$ increases.

Next, we analyze the case where $\lambda=4$ in Fig.~\ref{fig19} and Fig.~\ref{fig20}, comparing them with Fig.~\ref{fig17} and Fig.~\ref{fig18}. In Fig.~\ref{fig19}, we explore the variation of the Einstein ring radius $r_R$ with temperature $T$ for different chemical potentials $u=0.1$, $u=0.5$, and $u=1$, depicted in subfigures (a), (b), and (c) respectively, when $\lambda=4$. Similarly, the radius $r_R$ of the Einstein ring in units of $f$ is represented by discrete red dots, while the solid curves indicate geometric optics results. From Fig.~\ref{fig19}, we observe that when the chemical potential is small, such as $u=0.1$ and $u=0.5$, the ring radius does not exhibit monotonic behavior with increasing temperature $T$. However, when the chemical potential $u$ increases to $u=1$, we find that the radius of the ring decreases as the temperature $T$ increases, contrasting with the $\lambda=2$ case.

Similarly, in Fig.~\ref{fig20}, we study the changes in the Einstein ring radius $r_R$ with different chemical potentials $u$ for fixed temperatures $T=0.35$, $T=0.55$, and $T=0.95$, shown in subfigures (a), (b), and (c) respectively, when $\lambda=4$. The Einstein ring radius $r_R$ in units of $f$ is shown with discrete red dots, and the solid curves represent predictions from geometric optics. From Fig.~\ref{fig20}, it is observed that at $T=0.3$, the radius of the Einstein ring decreases slowly with increasing chemical potential $u$. As the temperature $T$ increases to $T=0.55$ and $T=0.95$, the ring radius shows minimal change with respect to the variation in chemical potential $u$, consistent with previous findings~\cite{Liu:2022cev}.

In addition, from  Fig.~\ref{fig18} and Fig.~\ref{fig20}, we know that at low temperature, the ring radius decreases as the chemical potential increases. While for enough high temperature, the ring radius changes little. This conclusion is consistent with that in Fig.~\ref{fig88} and Fig.~\ref{fig100}.   
From Fig.~\ref{fig17}, we know that for the small chemical potential, the ring radius decreases as the temperature increases while for the large chemical potential the ring radius increases as the temperature increases. This conclusion is consistent with that in Fig.~\ref{fig12}
and Fig.~\ref{fig14}.  
From Fig.~\ref{fig19}, we know that the ring radius decreases as the temperature increases for all the chemical potential, this conclusion is also consistent with that in Fig.~\ref{fig16}
and Fig.~\ref{fig160}.

These figures collectively show that the results obtained by wave optics via holography are consistent with those obtained by geometric optics, confirmed the validity of the holographic method to investigate the images of the black holes. {And we say more about the above comparison results. The Einstein radius calculated from geodesic analysis is shown by the blue curves. The curve seems to be consistent nearly with the Einstein radius of the image constructed from the response function in the wave optics (small differences may be due to numerical accuracy and wave effects). This indicates that the major contribution to the brightest ring in such image is originated by the “light rays” from the vicinity of the photon sphere, which are infinitely accumulated. The deviation of the Einstein radius from the geodesic prediction can be considered as some wave effects~\cite{Hashimoto:2018okj}. In the AdS cases, whether the geometrical optics can adapt to imaging of black holes is not so trivial even for a large value of $\omega$. As stated in Appendix B of~\cite{Hashimoto:2018okj}, the Eikonal approximation, which supports the geometrical optics, will inevitably break down near the AdS boundary, while we have given the source $J_{\mathcal{O}}$ and read the response \(\langle \mathcal{O} \rangle_{J_{\mathcal{O}}}\) on the AdS boundary. Our results based on the wave optics imply that the geometrical optics is qualitatively valid but gives a non-negligible deviation even for a large $\omega$.}

\section{Conclusions and discussions}
\label{4}
The Event Horizon Telescope (EHT) achieved a groundbreaking milestone in 2019 by capturing the first image of the shadow of the Supermassive Black Hole (SMBH) in the center of the $M87^{*}$ galaxy. This discovery opens up new avenues for constraining modified theories of gravity, as the shadow's size can be influenced by both new parameters introduced by alternative gravity models and the specific properties of the black hole solution, which differ from those predicted by General Relativity (GR).

In this study, we focus on investigating the Einstein ring of a charged black hole within Lorentz symmetry breaking massive gravity using the holographic method. We analyze phenomena such as photon rings, luminosity-deformed rings, or light points at various observation positions. Initially, we explored the influence of the parameter $\lambda$ in our gravity model on the response function, finding that different values of $\lambda$ exhibit nearly identical effects on the correlation function. Additionally, we investigated the impact of the frequency $\bar{\omega}$ on the Einstein ring, noting that higher frequencies tend to align more closely with geometric optics principles, thereby exerting less influence on the Einstein ring.

A critical aspect of our investigation focuses on the role of the chemical potential $u$ on the Einstein ring. Unlike previous studies that predominantly examined the effect of chemical potential on the ring radius at high temperatures, we extended our analysis to include low temperature scenarios. Our findings reveal that at low temperatures (as shown in Fig.~\ref{fig18}) and high temperatures (as shown in Fig.~\ref{fig20}), the radius of the photon ring decreases with increasing chemical potential $u$, consistent with prior research. Furthermore, we examined how the model parameter $\lambda$ impacts the radius of the Einstein ring. Specifically, for $\lambda=2$, the black hole characteristics resemble those of the RN-AdS black hole. Interestingly, for a chemical potential $u=1$, the relationship between the radius of the photon ring and temperature differs between $\lambda=2$ and $\lambda=4$. Conversely, for a smaller chemical potential such as $u=0.1$, the dependence of the Einstein ring on temperature remains consistent between $\lambda=2$ and $\lambda=4$.

{In fact, the results obtained by holography are similar to those obtained by studying Einstein rings through geometric optics, that is, the ring radius is the same, which is also emphasized in this paper. The difference is that through the holographic scheme, we find that the shadow shape is related to the wave source, such as the wave source frequency. Of course, the shadow image of any black hole can be compared with the results obtained by the EHT, and the parameters of the gravitational model can be restricted. And there is a lot of work in this direction. This paper does not focus on this problem, so it does not restrict the parameters of the black holes based on the EHT data, but in principle, it is feasible.}

In conclusion, our study underscores the pivotal role of holographic images in discerning the geometric nuances of different black holes within a fixed wave source and optical system setup. This research contributes to our understanding of how varying parameters in modified gravity models can alter observable phenomena such as the Einstein ring, thereby enriching our grasp of gravitational physics beyond the predictions of General Relativity.

\Acknowledgements{This work is supported  by the National
Natural Science Foundation of China (Grants No. 11675140, No. 11705005, and No. 12375043), and  Innovation and Development Joint  Foundation of Chongqing Natural Science  Foundation (Grant No. CSTB2022NSCQ-LZX0021) and Basic Research Project of Science and Technology Committee of Chongqing (Grant No. CSTB2023NSCQ-MSX0324).}

\InterestConflict{The authors declare that they have no conflict of interest.}








\end{multicols}
\end{document}